\documentclass[conference]{IEEEtran}
\usepackage{epigraph} 

\usepackage{tikz}
\usepackage{amsmath}
\usepackage{float}
\usepackage{hyperref}
\hypersetup{colorlinks,urlcolor=blue,citecolor=blue,breaklinks=true}
\usepackage{url}

\usepackage{breakurl}
\usepackage{cellspace}
\usepackage{booktabs,adjustbox}
\usepackage{tcolorbox}
\usepackage{bbm}
\newtcolorbox{mybox}{colback=black!5!white,colframe=black,bottomrule=.25mm,toprule=.25mm,leftrule=.25mm,rightrule=.25mm,left=.25mm,right=.25mm,top=-.25mm,bottom=-.25mm}

\usepackage{amsfonts}
\usepackage{mathtools} 
\usepackage[english]{babel}
\usepackage{amsmath,amssymb,amsthm}
\usepackage{tabularx}
\usepackage{algorithm}
\usepackage{algpseudocode}

\usepackage{multirow}
\usepackage{array}
\usepackage{caption}
\usepackage{listings}
\usepackage{diagbox}
\usepackage{tikz}
\usetikzlibrary{shapes,snakes}
\usepackage{subfloat}
\usepackage{subcaption}
\usepackage{empheq}
\usepackage{slashbox}

\usepackage{xcolor}
\usepackage{graphicx}
\usepackage{mhchem}
\usepackage{siunitx}
\usepackage{xspace}

\makeatletter

\def\addlegendimage{\pgfplots@addlegendimage}
\makeatother

\usepackage{pgfplots}
\usepackage{pgfplotstable}
\pgfplotstableset{
    /color cells/min/.initial=0,
    /color cells/max/.initial=1000,
    /color cells/textcolor/.initial=,
    color cells/.code={%
        \pgfqkeys{/color cells}{#1}%
        \pgfkeysalso{%
            postproc cell content/.code={%
                \begingroup
                \pgfkeysgetvalue{/pgfplots/table/@preprocessed cell content}\value
                \ifx\value\empty
                    \endgroup
                \else
                \pgfmathfloatparsenumber{\value}%
                \pgfmathfloattofixed{\pgfmathresult}%
                \let\value=\pgfmathresult
                \pgfplotscolormapaccess
                    [\pgfkeysvalueof{/color cells/min}:\pgfkeysvalueof{/color cells/max}]
                    {\value}
                    {\pgfkeysvalueof{/pgfplots/colormap name}}%
                \pgfkeysgetvalue{/pgfplots/table/@cell content}\typesetvalue
                \pgfkeysgetvalue{/color cells/textcolor}\textcolorvalue
                \toks0=\expandafter{\typesetvalue}%
                \xdef\temp{%
                    \noexpand\pgfkeysalso{%
                        @cell content={%
                            \noexpand\cellcolor[rgb]{\pgfmathresult}%
                            \noexpand\definecolor{mapped color}{rgb}{\pgfmathresult}%
                            \ifx\textcolorvalue\empty
                            \else
                                \noexpand\color{\textcolorvalue}%
                            \fi
                            \the\toks0 %
                        }%
                    }%
                }%
                \endgroup
                \temp
                \fi
            }%
        }%
    }
}
\pgfplotsset{compat=newest}
\usepgfplotslibrary{groupplots}
\usepgfplotslibrary{dateplot}
\pgfplotsset{every axis/.append style={
                    label style={font=\small},
                    tick label style={font=\small}  
                    }}

\usepackage{pifont}%
\usepackage{comment}

\usepackage[utf8]{inputenc}

\usepackage{colortbl}

\definecolor{auburn}{rgb}{0.43, 0.21, 0.1}
\definecolor{burgundy}{rgb}{0.5, 0.0, 0.13}

\newcommand{\tabincell}[2]{\begin{tabular}{@{}#1@{}}#2\end{tabular}}
\newcolumntype{P}[1]{>{\centering\arraybackslash}p{#1}}

\begin{document}

\date{}

\title{The Perils of Learning From Unlabeled Data:\\ Backdoor Attacks on Semi-supervised Learning}

\author{\normalsize Virat Shejwalkar$^*$,\ \  Lingjuan Lyu$^\dagger$,\ \ Amir Houmansadr$^*$\\
\normalsize {\normalsize $^*$University of Massachusetts Amherst\hspace{.5em}
$^\dagger$Sony AI}\\
\normalsize $^*$\{vshejwalkar, amir\}@cs.umass.edu, $^\dagger$Lingjuan.Lv@sony.com
}

\maketitle

\begin{abstract}
Semi-supervised machine learning  (SSL) is gaining popularity as it reduces the cost of training ML models. It does so by using very small amounts of (expensive, well-inspected) labeled data and large amounts of (cheap, non-inspected) unlabeled data. SSL has shown comparable or even superior performances compared to conventional fully-supervised ML techniques.

In this paper, we show that the key feature of SSL that it can learn from (non-inspected) unlabeled data exposes SSL to strong poisoning attacks. In fact, we argue that, due to its reliance on non-inspected unlabeled data, poisoning is a much more severe problem in SSL than in conventional~fully-supervised ML.

Specifically, we  design a \emph{backdoor poisoning attack} on SSL that can be conducted by a \emph{weak adversary} with  
no knowledge of target SSL pipeline. This is unlike prior poisoning attacks in fully-supervised settings that assume strong adversaries with practically-unrealistic capabilities. We show that by poisoning only 0.2\% of the unlabeled training data, our attack can cause misclassification of more than 80\% of test inputs (when they contain the adversary's backdoor trigger). Our attacks remain effective across twenty combinations of benchmark datasets and SSL algorithms, and even circumvent the state-of-the-art defenses against backdoor attacks. Our work raises significant concerns about the practical utility of existing SSL algorithms.

\end{abstract}

\section{Introduction}\label{intro}

The more training data we use to train machine learning (ML) models, the better they perform~\cite{deng2009imagenet,deng2009large}.
However, this makes the conventional \emph{fully-supervised} ML significantly challenging, as it requires large amounts of \emph{labeled} training data. Labeling is an expensive~\cite{culotta2005reducing} and error prone process~\cite{natarajan2013learning,li2017learning} that makes conventional ML prohibitively expensive in practice, especially with today's exploding training data sizes.

\emph{Semi-supervised learning (SSL)} addresses this major challenge by significantly reducing the need of the labeled training data: SSL uses a  combination of a \emph{small, high-quality and expensive labeled data} with a \emph{large, low-quality, and cheap  unlabeled data} to train its models. For instance, the  FixMatch~\cite{sohn2020fixmatch} SSL algorithm uses only \emph{40 labeled data} along with about 50,000 unlabeled data and achieves  90\% accuracy on CIFAR10.  
In contrast to fully-supervised algorithms, training an SSL algorithm involves two loss functions: a \emph{supervised loss} function (e.g., cross-entropy~\cite{murphy2012machine}) on labeled training data and an \emph{unsupervised loss} function (e.g., cross-entropy over pseudo-labels~\cite{lee2013pseudo}) on unlabeled training data. Different SSL algorithms primarily differ on how they compute their unsupervised losses. 

SSL is being explored extensively by both academia~\cite{zhang2021flexmatch,xie2020unsupervised,xie2020self} and industry~\cite{sohn2020fixmatch,sohn2020simple,berthelot2019mixmatch,berthelot2019remixmatch},  as  
recent SSL algorithms offer state-of-the-art performances comparable or even superior to those achieved by conventional supervised techniques\textemdash but with no need of large well-inspected labeled data. For instance, with less than 10\% of training data labeled, FixMatch~\cite{sohn2020fixmatch} and FlexMatch~\cite{zhang2021flexmatch} SSL algorithms outperform the fully-supervised algorithm.
This is because state-of-the-art SSL algorithms use (cheap, abundant) unlabeled data much more effectively than how fully-supervised algorithms use significantly larger (expensive, scarce) labeled data.

\noindent\textbf{\em Unlabeled data enables poisoning by weaker adversaries:} 
Multiple researches have demonstrated the threat that data poisoning attacks pose to fully-supervised learning~\cite{gu2017badnets,liu2017trojaning,saha2020hidden,sarkar2020facehack,zeng2022narcissus,turner2019label}. 
However, as the training data of fully-supervised models undergo an extensive and careful inspection, these attacks assume strong adversaries with the knowledge of model parameters~\cite{liu2017trojaning}, training data~\cite{turner2019label,biggio2012poisoning,munoz2017towards} or its distribution~\cite{zeng2022narcissus}, or the learning algorithm. Such strong adversaries are important to evaluate the worse-case security of a system, but are practically less relevant. 
On the other hand, the key feature of  SSL in real-world applications is that it  can leverage large amounts of\textemdash raw, non-inspected\textemdash unlabeled data, e.g., the data scraped from the Internet.
We argue that \textbf{the use of non-inspected data by SSL presents a significant threat to its security, as it allows even the most naive adversaries (with no knowledge of training algorithm, labeled data, etc.) to poison SSL models} by simply fabricating malicious unlabeled data. 
Unfortunately, this severe threat remains largely unexplored in the SSL literature.

To address this gap, in this paper, we take the first step towards thoroughly understanding this threat by studying the possibility of \emph{backdoor attacks} against SSL in real-world settings.
A backdoor attack aims to install a \emph{backdoor function} in the \emph{target model}, such that the \emph{backdoored target model} will \emph{misclassify} any test input to the \emph{adversary chosen target class} when patched with a specific \emph{backdoor trigger}, but will correctly classify the input without the trigger.

\noindent\textbf{\em Existing backdoor attacks fail on SSL:}
There exist numerous backdoor attacks in prior literature, however, except one attack\textemdash \emph{DeHiB}~\cite{yan2021dehib}, \emph{all of the prior attacks consider  fully-supervised settings}. 
Our preliminary evaluations show that all of the existing state-of-the-art attacks, including DeHiB, completely fail against SSL under our realistic threat model (Section~\ref{threat_model}).
Hence, to learn from these failures, we first systematically evaluate five backdoor attacks from three categories against five state-of-the-art SSL algorithms, under our practical, unlabeled data poisoning threat model. We adaptively choose the attack categories based on the specific lesson we learn from evaluating the prior category.

Our systematic evaluation leads to the following \textbf{three major lessons} that not only guide our attack design, but are generally applicable to any (future) backdoor attacks against SSL: (1) \emph{Backdoor attacks on SSL should be clean-label style attacks}, i.e., poisoning data should be selected from the target class distribution; (2) \emph{Backdoor trigger patterns should span/cover the entire poisoning sample}, to circumvent strong augmentations, e.g., cutout~\cite{devries2017improved}, that all modern SSL algorithms use; (3) \emph{Backdoor trigger patterns should be resistant to noise and with repetitive patterns} to withstand large amounts of random noises which were added to training data via the strong augmentation in SSL.

\noindent\textbf{\em Our SSL-tailored backdoor method:}
The high-level intuition behind our backdoor attack is as follows.
All modern semi-supervised algorithms learn via a self-feedback mechanism, called \emph{pseudo-labeling}, i.e., the prediction $\tilde{y}$ on an unlabeled sample $\mathbf{x}$ has high confidence, they use ($\mathbf{x},\tilde{y}$) as a labeled sample for further training. Following our first lesson, we exploit this pseudo-labeling and design a \emph{clean-label} attack that poisons samples $\mathbf{x}$ only from the distribution of the target class $y^t$. Our attack patiently \emph{waits} for the target model to correctly label a poisoning sample $(\mathbf{x} + T)$ as $y^t$, where $T$ is our backdoor trigger. And then, as the model trains further on ($(\mathbf{x} + T), y^t$), our attack \emph{forces the model} to associate features of our simple trigger $T$, instead of the complex features of $\mathbf{x}$, with the target class, which effectively installs the backdoor function in the target model.

Note that,  we consider the most challenging setting for designing attacks with the least capable and knowledgeable data poisoning adversary. Generally, trigger generation for data poisoning backdoor attacks is formalized as a bi-level optimization problem~\cite{munoz2017towards}, however such attacks are well-known to be very expensive, and yet ineffective~\cite{munoz2017towards,shejwalkar2022back}.
Instead, we design a static, repetitive grid pattern for our backdoor trigger (Figure~\ref{fig:our_attack_trigger}), as guided by lessons 2 and 3. In summary, we sample few data from target class, patch them with our trigger and inject into unlabeled training data.

\noindent\textbf{\em Evaluations:}
We demonstrate the strength of our attack via an extensive evaluation against five state-of-the-art SSL algorithms and a fully-supervised algorithm, using four benchmark image classification tasks, that SSL literature commonly uses. 
We note that our attack significantly outperforms the prior attacks from both SSL and full-supervised literature.

For the most combinations of algorithm and dataset, \textbf{our attacks achieve high attack success rates (ASRs) ($>$80\%), while poisoning just  0.2\% of entire training data}. Comparatively, DeHib uses 20$\times$ more poisoning data and achieves 0\% ASRs. 
ASR measures the \% of \emph{test inputs from non-target classes} that the backdoored model classifies to the target class when patched with backdoor trigger. 
For instance, our attacks have more than 90\% ASR against CIFAR100 and more than 80\% ASR against CIFAR10. For SVHN and STL10, our attack has more than 80\% ASR with two exceptions each.
Through a systematic experiment design in Section~\ref{exp:attack_dynamics}, we show that our intuition aligns with the dynamics of our patient attacks and justify their strength.
\textbf{Our attack is highly stealthy}, as (1) it minimally perturbs the poisoning data and (2) it produces backdoored models which have high accuracy (close to non-backdoored models) on non-backdoored test inputs.

Next, our comprehensive ablation study (Section~\ref{exp:ablation}) shows the high efficacy of our attacks when varying three major parameters of our setting: size of labeled data, backdoor target class, and size of poisoning data. 
Finally, we show that \textbf{our attacks remain highly effective even when SSL is paired individually with five state-of-the-art defenses} against backdoor attacks that are agnostic to learning algorithms.

\noindent\textbf{\em Summary of contributions:}
\vspace*{-.2em}

\noindent\textbf{(1)} We perform the first thorough study of backdoor attacks on semi-supervised learning (SSL), and show that it is highly susceptible to backdoor poisoning, under realistic unlabeled data poisoning threat models.
\vspace*{-.3em}

\noindent\textbf{(2)} We systematically evaluate existing backdoor attacks from fully-supervised setting on SSL and provide concrete lessons to design stronger backdoor attacks against SSL.
\vspace*{-.3em}

\noindent\textbf{(3)} Based on the lessons, we design the first effective backdoor attack against SSL that achieves high ($>$80\%) ASRs by poisoning just 0.2\% of entire training data.
\vspace*{-.3em}

\noindent\textbf{(4)} We show that existing learning-algorithm-agnostic defenses are insufficient to defend SSL against backdoor attacks.

\section{Preliminaries and Related Work}\label{prelims}
In this section, we provide the preliminaries and important related works required to understand the rest of the paper.

\subsection{Semi-supervised Learning (SSL)}\label{prelims:semi_supervised}
The objective of machine learning (ML) for classification task is to train a classifier (e.g., neural network) with parameters $\theta$ and learn function $f_{\theta}:\mathcal{X} \mapsto \mathcal{Y}$ to predict label $y\in\mathcal{Y}$ for input $\mathbf{x}\in\mathcal{X}$. Here, $\mathcal{X}\in\mathbb{R}^d$ is the input feature vector space and $\mathcal{Y}\in\mathbb{R}^k$ is the output space. Parameters $\theta$ are trained using empirical risk minimization (ERM) to minimize certain empirical loss function $\ell_{(\mathbf{x},y)\in D}(f_{\theta}, (\mathbf{x},y))$, where $D$ is the training data and $D\subset \mathcal{X}\times\mathcal{Y}$.
Traditionally, ML uses \emph{fully-supervised} learning algorithms that use only completely labeled data, $D^l$. However, labeling is a manual, expensive, and error-prone process~\cite{li2017learning}, as it requires human intervention. Consequently, with continuously increasing training data sizes, labeling cost for ML can become prohibitively high in practice. 

\emph{Semi-supervised learning (SSL)} addresses this major challenge by reducing the dependence of ML on labeled data. SSL proposes to learn ML models using both labeled $D^l$ and unlabeled data $D^u$. $D^l$ and $D^u$ may or may not have the same distribution, but sizes of labeled data are significantly smaller than that of unlabeled data, i.e., $|D^l|\ll |D^u|$.
A typical SSL loss function is a convex combination of a loss on $D^l$, denoted by $\mathcal{L}_l$, and a loss on $D^u$, denoted by $\mathcal{L}_u$: $\mathcal{L}_{ss} = \mathcal{L}_l + \lambda\mathcal{L}_u$. $\mathcal{L}_l$ is generally the standard cross-entropy loss due to its high performances. But, $\mathcal{L}_u$ varies across different SSL algorithms; we will discuss these shortly.

Traditionally, semi-supervised learning has been largely ineffective, but in the past couple of years, SSL has significantly improved, especially after the invention of MixMatch~\cite{berthelot2019mixmatch}. Significant performance gains of MixMatch are because it combines various data augmentation techniques with various prior SSL algorithms, including \emph{pseudo-labeling}~\cite{lee2013pseudo}, entropy minimization~\cite{grandvalet2004semi}, and \emph{consistency regularization}~\cite{denton2016semi,sajjadi2016regularization,laine2016temporal}. 
Hence, in this work we consider the state-of-the-art algorithms that use pseudo-labeling and {consistency regularization}. We briefly describe these two techniques as in~\cite{sohn2020fixmatch} followed by the semi-supervised algorithms we consider in this work.

\subsubsection{Pseudo-labeling} Pseudo-labeling uses the current model to obtain artificial \emph{pseudo-labels} for the unlabeled data. There are various ways in which model's outputs can be used to train itself, e.g., MixMatch and ReMixMatch compute model's predictions on an unlabeled sample and then use entropy minimization~\cite{grandvalet2004semi} to sharpen the prediction.
But, pseudo-labeling specifically refers to the use of ``hard'' labels (i.e., the \texttt{argmax} of the model's output) and only retains the labels whose largest class probability (confidence) is above a pre-defined threshold. Assume $q_b=f_\theta(y|u_b)$ are the predictions of current model $f_\theta$ on the batch $u_b$ of unlabeled data. Then pseudo-labeling loss can be formalized as follows:
\begin{align}\label{eq:pseudo_label}
\frac{1}{|u_b|}\sum^{|u_b|}_{b=1} \mathbbm{1}(\texttt{max}(q_b) \geq \tau) H(\hat{q_b}, q_b)
\end{align}
where $\hat{q_b} = \texttt{argmax}(q_b)$, $H(.)$ is the cross-entropy loss function, and $\tau$ is the threshold parameter. Note that the use of hard labels is similar to entropy minimization in MixMatch or ReMixMatch, where they use a temperature parameter to sharpen the model predictions to have low entropy (high confidence).

\subsubsection{Consistency regularization} Consistency regularization is commonly used in modern semi-supervised learning algorithms. It is based on the intuition that the model should output similar predictions when input with the perturbed versions of the same image. The idea was first proposed in~\cite{bachman2014learning}. Semi-supervised learning algorithms use consistency regularization on unlabeled data as follows. They use a stochastic augmentation mechanism $a(\mathbf{x}_u)$ to produce perturbed versions of an unlabeled sample $\mathbf{x}_u$ and then force the model to have similar outputs on these versions using the following loss:
\begin{align}\label{eq:consistency_reg}
    \sum^{|u_b|}_{b=1} \Vert f_\theta(y|a(u_b)) - f_\theta(y|a(u_b)) \Vert^2_2
\end{align}
where, $a(.)$ is a stochastic function, hence it produces different output every time it is applied to a batch $u_b$ of unlabeled data. Consequently, the two terms in~\eqref{eq:consistency_reg} have different values.
Next, we briefly describe the semi-supervised learning algorithms we consider in this work. For detailed description of the algorithms, please refer to the original works.

\noindent\textbf{(1)} \emph{MixMatch}~\cite{berthelot2019mixmatch} combines various prior semi-supervised learning techniques. For an unlabeled sample, MixMatch generates $K$ weakly augmented versions of the unlabeled sample, computes outputs of the current model $f_\theta$ for the $K$ versions, averages them, and sharpens the average prediction by raising all its probabilities by a power of 1/temperature and re-normalizing; it uses the sharpened prediction as the label of the unlabeled sample. Finally, it uses mixup regularization~\cite{zhang2018mixup} on the combination of labeled and unlabeled data and trains the model using cross-entropy loss.

\noindent\textbf{(2)}
\emph{Unsupervised data augmentation (UDA)}~\cite{xie2020unsupervised} shows significant improvements in semi-supervised performances by just replacing the simple weak augmentations of MixMatch with a strong augmentation called Randaugment~\cite{cubuk2020randaugment}. In an iteration, Randaugment randomly selects a few augmentations from a large set of augmentations and applies them to images.

\noindent\textbf{(1)}
\emph{ReMixMatch}~\cite{berthelot2019mixmatch} builds on MixMatch by making multiple modifications, including 1) it replaces the simple weak augmentation in MixMatch with Autoagument~\cite{cubuk2018autoaugment}, 2) it uses augmentation anchoring to improve consistency regularization, i.e., it uses the prediction on a weakly augmented version of unlabeled sample as the target prediction for a strongly augmented version of the unlabeled sample, and 3) it uses distribution alignment, i.e., it normalizes the new model predictions on unlabeled data using the running average of model predictions on unlabeled data. This significantly boosts the performance of resulting model.

\noindent\textbf{(4)} \emph{FixMatch}~\cite{sohn2020fixmatch} simplifies the complex ReMixMatch algorithm by proposing to use a combination of Pseudo-labeling and consistency regularization based on augmentation anchoring (discussed above). FixMatch significantly improves semi-supervised algorithms, especially in the low labeled data regimes.

\noindent\textbf{(5)} 
\emph{FlexMatch}~\cite{zhang2021flexmatch} proposes curriculum pseudo labelling (CPL) approach to leverage unlabeled data according to model's learning status. The main idea behind CPL is to flexibly adjust the thresholds used for pseudo-labeling for different classes at each training iteration in order to select more information unlabeled data and their pseudo-labels. CPL can be combined with other algorithms, e.g., UDA.

\subsection{Backdoor Attacks}\label{prelims:backdoor_attacks}
A \emph{backdoor adversary} aims to implant a \emph{backdoor functionality} into a \emph{target} model. That is, given an input $(\mathbf{x},y^*)$ with true label $y^*$, the \emph{backdoored target model} $f^b_\theta$ should output an adversary-desired \emph{backdoor target label} $y^t$ for the input patched with a pre-specified \emph{backdoor trigger} $T$, but it should output the correct label for the benign input, i.e., $f^b_\theta(\mathbf{x}+T)\mapsto y^t$ and $f^b_\theta(\mathbf{x})\mapsto y^*$.
There are two major types of backdoor attacks: \emph{dirty-label} and \emph{clean-label} backdoors. 

\subsubsection{Dirty-label backdoor attacks~\cite{gu2019badnets,chen2017targeted,sarkar2020facehack,zeng2021rethinking,nguyen2020input,li2021invisible}} These attacks poison both the features $\mathbf{x}$ and labels $y^*$ of benign, labeled data to obtain poisoning data $D^p$. They first select some benign data $(X, Y^{\backslash t})$ from non-target classes of the original training data $D$, patch the backdoor trigger to $X$: $X^p\leftarrow X+T$, and set labels of $X^p$ to $y^t$ to obtain $D\leftarrow D \cup D^p=(X^p, y^t)$. Training $f_\theta$ on such $D$ makes the model \emph{associate} the trigger with the target label, i.e., $f^b_\theta(T)\mapsto y^t$, as this association is much easier to learn than learning to associate original $X$ to $Y$. 
However, in many practical scenarios, a trusted third party conducts the data labeling and inspection, and can easily remove such mislabeled data. Our work focuses on the semi-supervised learning where such inspection is much easier compared to the supervised learning due to very small sizes of labeled data, hence we only poison the unlabeled training data.

\subsubsection{Clean-label backdoor attacks~\cite{turner2019label,zeng2022narcissus,zhong2020backdoor}} These attacks poison only the features $X$ of benign data. They add imperceptible $T$ to $X$ such that the poisoned features $X^p$ appear to be from the respective true classes to a human. Clean-label attacks can be further divided into two categories~\cite{zeng2022narcissus} based on the true classes of $X$ that they poison: \emph{feature-collision} attacks and \emph{target-class} attacks.

\noindent\emph{\hspace{.5cm}\textbf{(a)} Feature-collision backdoor attacks}, e.g., HTBA~\cite{saha2020hidden} and SAA~\cite{souri2021sleeper}, insert triggers indirectly. They try to match the feature space/gradients between the target class samples and non-target-class samples patched with the trigger, thus, mimicking the effects of non-target-class poisoning.
By doing this, the decision boundary will place these two points in proximity in the feature space, and as a result, any input with the trigger will likely be classified into the target class. However, these attacks can backdoor just one sample at a time, and hence, are computationally inefficient at backdooring large portions of test inputs.

\noindent\emph{\hspace{.5cm}\textbf{(b)} Target-class backdoor attacks} select $X$ to poison from the target class $y^t$. For instance, label-consistent (LC) backdoor attacks~\cite{turner2019label} select a few $X$ from $y^t$ and manipulate $X$ to make their original features harder to learn. Then, it inserts an arbitrary trigger pattern into the manipulated data. They either use GANs or adversarial samples to manipulate $X$ to obtain difficult-to-learn $X'$ and then add $T$ to get $X^p$, i.e., $X^p\leftarrow X' + T$. LC attack requires in-distribution data from both target and non-target classes in order to train adversarial example generator. Narcissus~\cite{zeng2022narcissus} attack addresses the above issues as it requires in-distribution data only from the target class, $y^t$. For clarity of presentation, we discuss more details of the Narcissus attack in Section~\ref{method:lessons3} and evaluate them against semi-supervised learning.

\subsubsection{Backdoor attacks on semi-supervised learning} 
So far, most of the backdoor attack literature has focused on the fully-supervised settings. Only one work by Zhicong et al.~\cite{yan2021dehib} study backdoor attacks against semi-supervised learning. For clarity of presentation, we discuss the details of this attack in Section~\ref{method:lessons1}, where we demonstrate and justify why this attack fails to backdoor semi-supervised learning.

\section{Threat Model}\label{threat_model}

Below, we discuss the threat model of our backdoor attacks in terms of the adversary's goal, knowledge, and capabilities.
We consider a setting where a victim model trainer collects data from multiple, potentially untrusted sources to train a ML model for a classification task with $C$ classes. 
Below, $[I]$ denotes the set of all non-zero positive integers $\leq I$.

\subsection{Adversary's goal} 

We consider a \emph{backdoor adversary} who aims to install a \emph{backdoor function} in the victim's ML model, called \emph{target model}. We denote the function of a \emph{benign model}, i.e., without any backdoor by $f_\theta$ and that of a \emph{backdoored (target) model} by $f^b_\theta$. Our adversary's goal is two-fold.

\subsubsection{Backdoor goal} 
The backdoor adversary selects a \emph{backdoor target class} $y^t \in[C]$. To mount an effective backdoor attack, the backdoor goal requires the backdoored model to \emph{incorrectly classify all the test inputs from non-target classes to the target class, when they are patched with a pre-specified backdoor trigger}, $T$. More formally, $f^b_\theta(\mathbf{x}+T)\mapsto y^t \ \forall \ (\mathbf{x},y^*) \text{ where } y^*\in [C]\backslash y^t$ is the true class of $\mathbf{x}$.

\subsubsection{Stealth goal}
In order to mount a stealthy backdoor attack, the stealth goal requires the backdoored model, $f^b_\theta$, to retain all the benign functionalities similar to the benign model, $f_\theta$. Specifically, $f^b_\theta$ should correctly classify all the benign test inputs from all the classes without the backdoor trigger. 
Formally, $f^b_\theta(\mathbf{x}) = f_\theta(\mathbf{x}) \mapsto y^* \  \forall \ (\mathbf{x},y^*) \text{ where } y^*\in [C]$.

\subsection{Adversary's knowledge}
As discussed in Section~\ref{intro}, we consider the most naive, real-world adversary with minimum knowledge of the semi-supervised learning (SSL) pipeline.
We assume that the adversary has no knowledge of the data except the specific classes of the classification task.
Next, the adversary knows the details of the target SSL algorithm, but do not know model architecture, e.g., ResNet or VGG, i.e., \emph{our attacks are model architecture agnostic}.
Finally, we assume that the adversary does not know the distribution of the unlabeled and labeled training data, or posses any data from the true distribution.

\subsection{Adversary's capabilities}

We discuss the adversary's capabilities based on their ability to manipulate the training pipeline and training data.

\subsubsection{Manipulating training pipeline} 
There are three types of poisoning attacks based on the part of SSL training pipeline that the adversary can manipulate: \emph{data poisoning}, \emph{code poisoning}, and \emph{model poisoning}.
The \emph{model poisoning}~\cite{bagdasaryan2018how,liu2017trojaning,shejwalkar2021manipulating} adversary is the strongest adversary who directly manipulates the model parameters. Such poisoning requires highly privileged accesses to the training platforms, which is impractical in many real-world settings, e.g., for popular ML platforms like Amazon AWS~\cite{shejwalkar2022back}. The \emph{code poisoning}~\cite{bagdasaryan2021blind,georgiev2012most} adversary poisons the code of the target algorithms and also requires appropriate permissions to change the code along with a thorough knowledge of the learning pipeline.

Finally, the \emph{data poisoning}~\cite{biggio2011bagging,munoz2017towards} adversary can only manipulate the training data of the target model. Due to its naivety, it is also the most practical adversary who can be any data owner willing to contribute data to SSL. Data poisoning is a severe threat because it is easy to deploy~\cite{shejwalkar2022back}, even against sophisticated ML platforms, e.g., Amazon AWS, Google cloud and Microsoft Azure, with state-of-the-art software security. Hence, \emph{we consider the data poisoning adversary in this work}.

\subsubsection{Manipulating training data}
We discuss manipulation of training data separately from that of training pipeline, especially because SSL uses two types of datasets.
SSL bootstraps knowledge from a small labeled dataset $D^l$, hence $D^l$ is of very high quality and is well-inspected. Therefore, we argue that an adversary who poisons or has access to $D^l$~\cite{yan2021dehib} is not practically relevant. On the other hand, a salient feature of semi-supervised learning is that the model trainer needs not to inspect its unlabeled data $D^u$ at all (Section~\ref{prelims:semi_supervised}), and hence, $D^u$ can be easily poisoned in practice. Hence, we assume that \emph{our adversary can poison only the unlabeled training data}.

\begin{table*}
\caption{Left-most column shows types of backdoor attacks based on specific characteristics, middle column lists existing attacks of each type. Right-most column presents lessons we learn from evaluating one/two representative attacks (in bold) of each type.}\label{tab:lessons}
\centering
\setlength{\extrarowheight}{0.01cm}
\hspace*{-1em}
\begin{tabular} {|c|P{5cm}|P{5.5cm}|}
\hline
  {\bf Attack characteristic/ type} & {\bf Existing attacks of given type} & {\bf Lesson from evaluations} \\ \hline
  {Dirty label} & {\bf DeHiB}~\cite{yan2021dehib}, {\bf DL-Badnets}~\cite{gu2017badnets}, DL-Blend~\cite{chen2017targeted}, Facehack~\cite{sarkar2020facehack} & \em Attack should be a clean-label attack, i.e., poisoning samples should be from backdoor target class. \\ \hline
  Clean-label small trigger & {\bf CL-Badnets}~\cite{zeng2022narcissus}, CL-Blend~\cite{chen2017targeted} & \em Trigger should span the entire sample/image to avoid cropping/covering by strong augmentations. \\ \hline
  Clean-label adversarial samples & {\bf Narcissus}~\cite{zeng2022narcissus}, {\bf Label consistent}~\cite{turner2019label}, {\bf non-repeating trigger patterns}, HTBA~\cite{saha2020hidden}, SAA~\cite{souri2021sleeper}, Embedding~\cite{zhong2020backdoor} & \em Trigger should be noise-resistant and its pattern should be repetitive so that even a part of trigger can install a backdoor. \\ \hline
\end{tabular}
\vspace*{-1.5em}
\end{table*}

\section{Our Attack Methodology}\label{attack_method}

In this section, we discuss our backdoor attack methodology tailored to the unlabeled data poisoning threat model (Section~\ref{threat_model}). In Section~\ref{method:systematic_eval}, we present the first systematic evaluation of existing backdoor attacks in semi-supervised learning (SSL) settings and provide three major lessons.  Next, we give the intuition behind our backdoor attack (Section~\ref{method:intuition}), and finally we detail our attack method in Section~\ref{method:our_trigger}.

\begin{figure}
\vspace*{-.75cm}
\begin{subfigure}{\columnwidth}
  \centering
  \hspace*{-1em}
  \includegraphics[scale=.57]{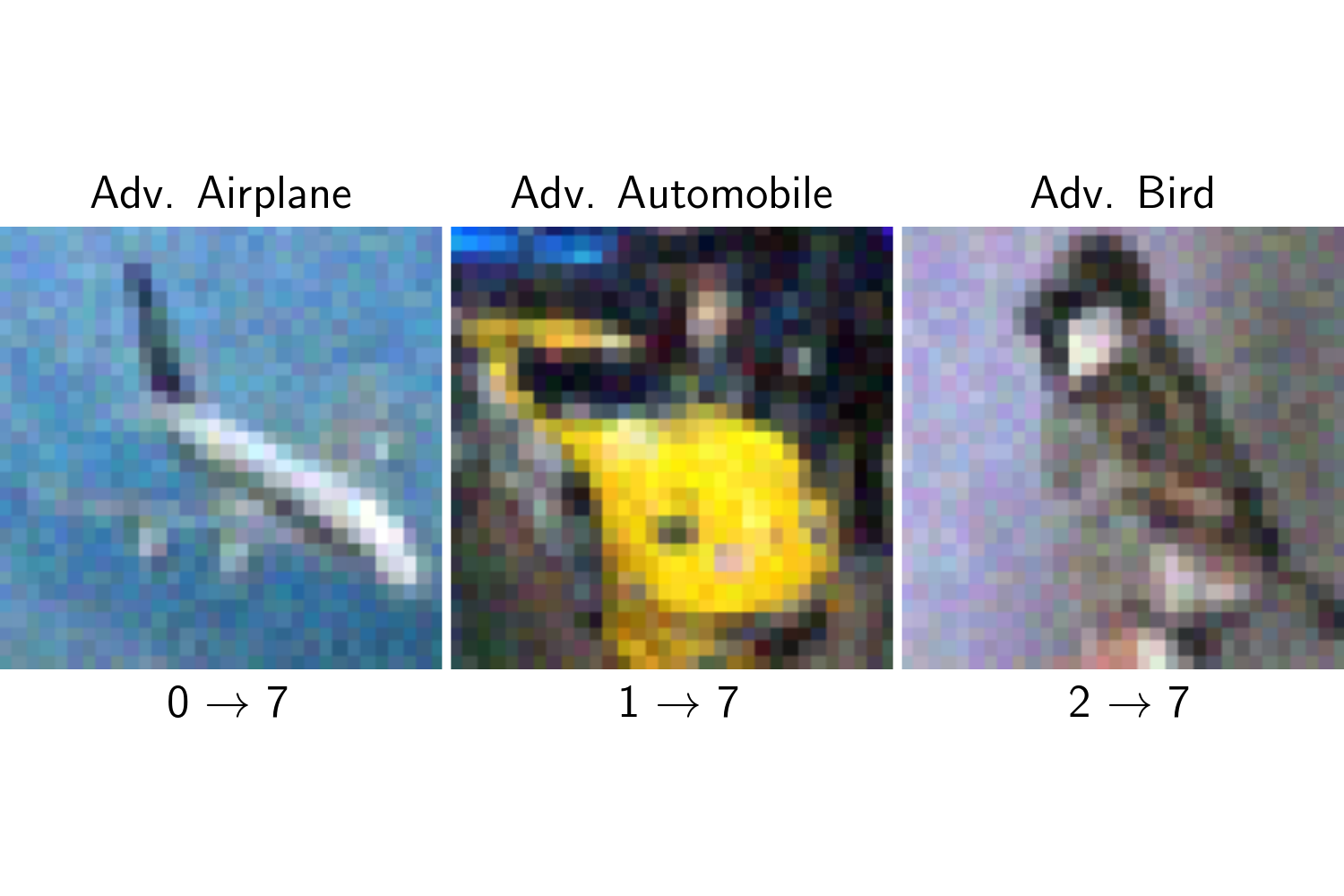}
  \vspace*{-1.2cm}
  \caption{\small Before random-crop augmentation}
  \label{fig:failure_dehib_before_aug}
\end{subfigure}%
\vspace*{-3.75em}
\newline
\begin{subfigure}{\columnwidth}
  \centering
  \hspace*{-1em}
  \includegraphics[scale=.57]{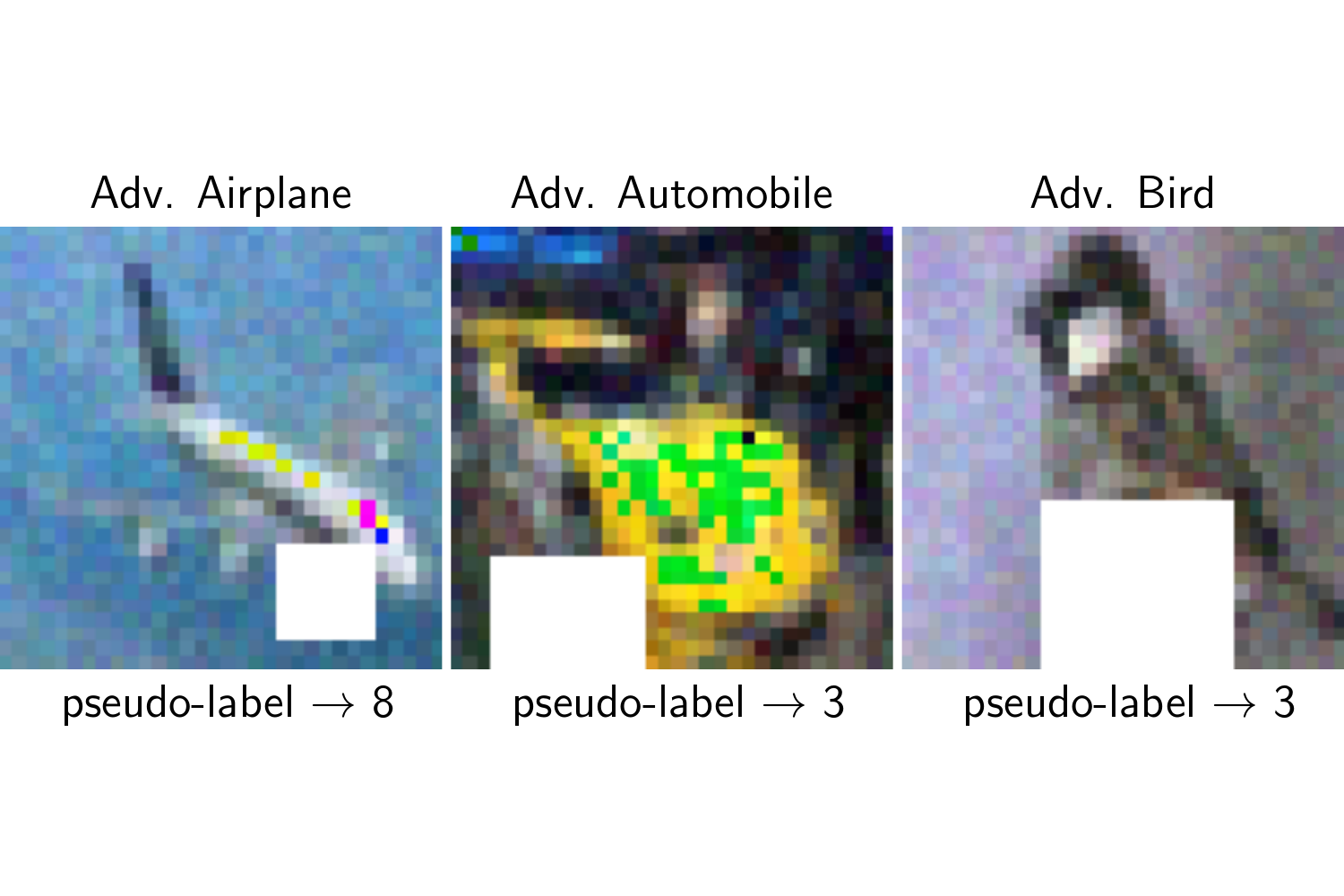}
  \vspace*{-1.2cm}
  \caption{After random-crop augmentation}
  \label{fig:failure_dehib_after_aug}
\end{subfigure}%
\vspace*{-1.2em}
\caption{DeHiB~\cite{yan2021dehib} fails because it cannot obtain the target class as pseudo-labels for its poisoning data.}
\label{fig:failure_dehib}
\vspace*{-2em}
\end{figure}

\begin{table}
\caption{Impacts of existing backdoor attacks (Section~\ref{prelims:backdoor_attacks}) on various semi-supervised algorithms for CIFAR10 data. We poison 0.2\% (100 samples) of all the training data. DeHib$^*$ is the original attack with the knowledge of labeled training data $D^l$ while DeHib is the attack without the knowledge of $D^l$. }\label{tab:existing_attacks}
\centering
\hspace*{-1em}
\begin{tabular} {|c|c|c|c|c|c|}
  \hline
  \multirow{2}{*}{Algorithm} & {DeHiB$^*$} & {DeHiB} & {CL-Badnets} & {LC} & {Narcissus} \\ %
  & ASR (\%) & ASR (\%) & ASR (\%) & ASR (\%) & ASR (\%)\\ \hline
  
  Mixmatch~\cite{berthelot2019mixmatch} & 22.0 & 1.0 & 9.1 & 1.1 & 2.2 \\ \hline
  Remixmatch~\cite{berthelot2019remixmatch} & 10.9 & 0.9 & 0.0  & 0.0 & 0.0 \\ \hline
  UDA~\cite{xie2020unsupervised} & 21.2 & 1.2 & 5.1 & 0.0 & 0.0 \\ \hline
  Fixmatch~\cite{sohn2020fixmatch} & 35.8 & 0.9 & 10.2 & 0.1 & 1.3 \\ \hline
  Flexmatch~\cite{zhang2021flexmatch} & 16.9 & 1.2 & 9.1 & 0.1 & 1.1 \\ \hline
\end{tabular}
\vspace*{-1.5em}
\end{table}

\subsection{Systematic evaluation of existing backdoor attacks}\label{method:systematic_eval}

Previous works have proposed numerous backdoor attacks under different threat models. But all works, except DeHiB~\cite{yan2021dehib}, consider fully-supervised setting. 
Hence, we first present a systematic evaluation of existing state-of-the-art backdoor attacks and explain why they fail in SSL settings. Based on our evaluations, we provide three major lessons that are fundamental to our attack design and generally apply to any (future) backdoor attacks against semi-supervised learning.

We start our evaluations from DeHiB~\cite{yan2021dehib}, the only existing backdoor attack on semi-supervised learning, and based on the lessons learned from this evaluation, we chose the next type of attacks to evaluate.
As we see from Table~\ref{tab:lessons}, each of our lessons applies to multiple backdoor attacks of a specific type and characteristics. However, for conciseness, we evaluate one or two representative attacks from each type and provide lesson/s that are useful in designing stronger attacks.

\subsubsection{Attacks should be clean-label attacks}\label{method:lessons1}

We first evaluate \emph{Deep hidden backdoor} (DeHiB)~\cite{yan2021dehib} attack.
DeHiB poisons only the unlabeled data, $D^u$, but it assumes a strong, unrealistic adversary who can access the labeled data, $D^l$. It first samples some data $(X,Y)$ from both target, $y^t$, and non-target, $y^{\backslash t}$, classes. Then it uses a model trained on $D^l$ to add universal adversarial perturbation $\mathcal{P}_t$ to $X$ such that the perturbed data $X+\mathcal{P}_t\mapsto X^p$ is classified as $y^t$; as we only poison $D^u$, we denote poisoning data by $X^p$. Finally, it adds a static trigger $T$ to the perturbed data $X^p$. Intuition behind DeHiB is that, due to $\mathcal{P}_t$, SSL algorithm will assign target class $y^t$ as pseudo-labels to all $X^p$ and force the target model to associate static trigger $T$ to $y^t$ and ignore original features $X$.

\emph{Why does DeHiB fail?} Recall from Section~\ref{prelims:semi_supervised} that all of state-of-the-art SSL algorithms use various strong augmentations, including, cutout~\cite{devries2017improved}, adding various types of hue~\cite{shorten2019survey}, horizontal/vertical shifts~\cite{taylor2018improving}, etc. Next, note that adversarial perturbations are sensitive to noises~\cite{athalye2018synthesizing}, i.e., even moderate changes in the perturbations render them ineffective. Hence, in presence of strong augmentations, adversarial perturbations fail to obtain the backdoor target class $y^t$ as the pseudo-labels for $X^p$ of DeHiB as shown in Figure~\ref{fig:failure_dehib}. Hence, the very fundamental requirement of DeHiB does not hold in SSL and leads to its failure. 
The original DeHiB work reports slightly better results, because it assumes access to $D^l$, which our threat model does not allow. 
Hence, we use randomly sampled data of size $|D^l|$ from entire CIFAR10 data to obtain DeHiB's $\mathcal{P}_t$. 

To summarize, adversarial perturbations are sensitive to noises. Hence, using adversarial samples from non-target classes as poisoning samples cannot guarantee the desired pseudo-labeling to $y^t$. Effectively, such attack tries to train the model to associate the trigger pattern $T$ with multiple labels, and hence, fails to inject the backdoor functionality. For the same reason, \emph{we also observed that any dirty-label static trigger attacks completely fail against SSL}. Hence, backdoor attacks on SSL should be clean-label attacks, i.e., use poisoning samples $X^p$ from $y^t$, and leverage benign features of $X^p$ to obtain desired pseudo-labels $y^t$ for them. 

\begin{mybox}
\textbf{Lesson-\hyperref[method:lessons1]{1}}: Backdoor attacks on semi-supervised learning should be clean-label style attacks, which sample their poisoning samples from the backdoor target class.
\end{mybox}

\begin{figure}
\vspace*{-.75cm}
\begin{subfigure}{\columnwidth}
  \centering
  \hspace*{-1em}
  \includegraphics[scale=.57]{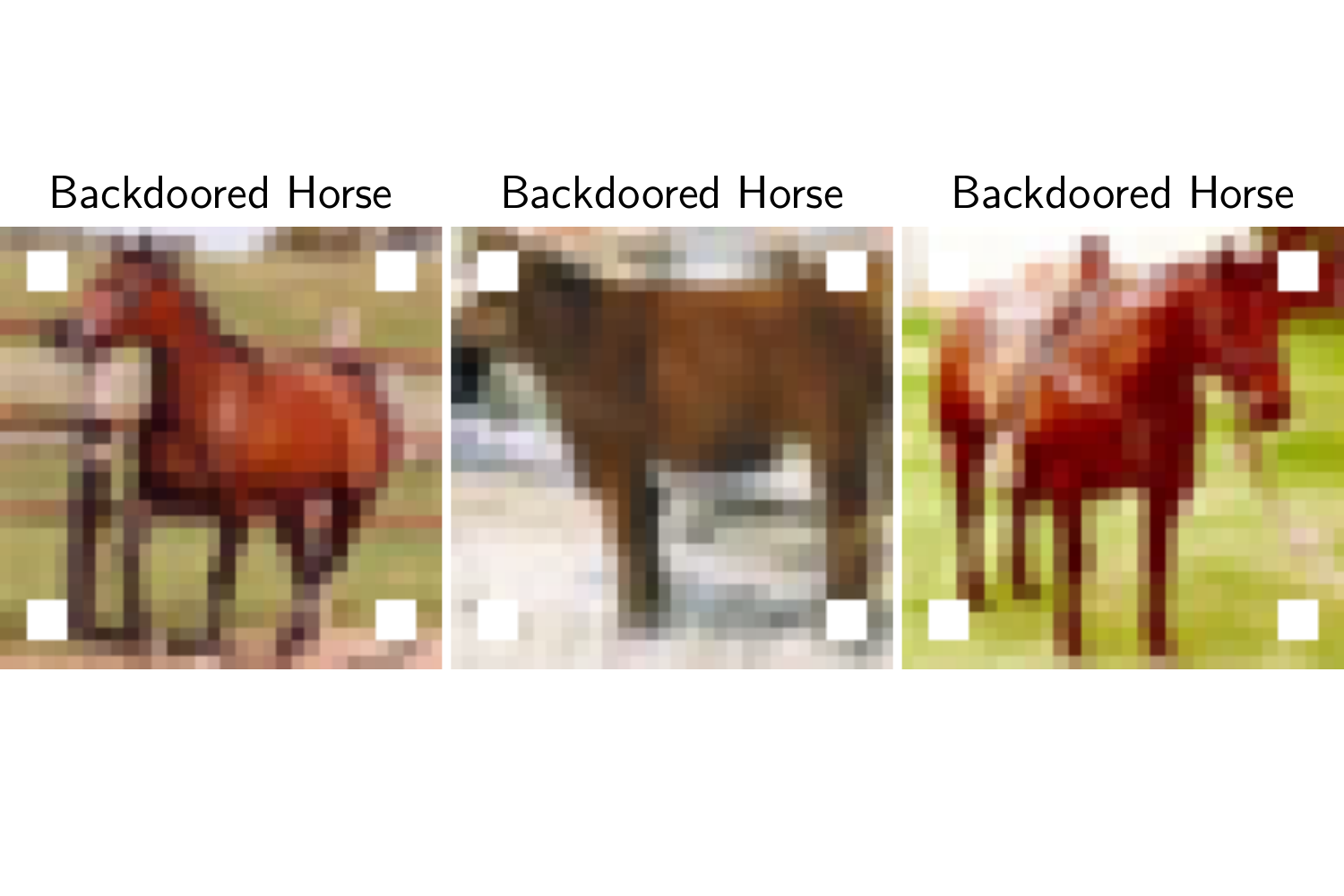}
  \vspace*{-1.5cm}
  \caption{\small Before random-crop augmentation}
  \label{fig:failure_clbadnets_before_aug}
\end{subfigure}%
\vspace*{-3.75em}
\newline
\begin{subfigure}{\columnwidth}
  \centering
  \hspace*{-1em}
  \includegraphics[scale=.57]{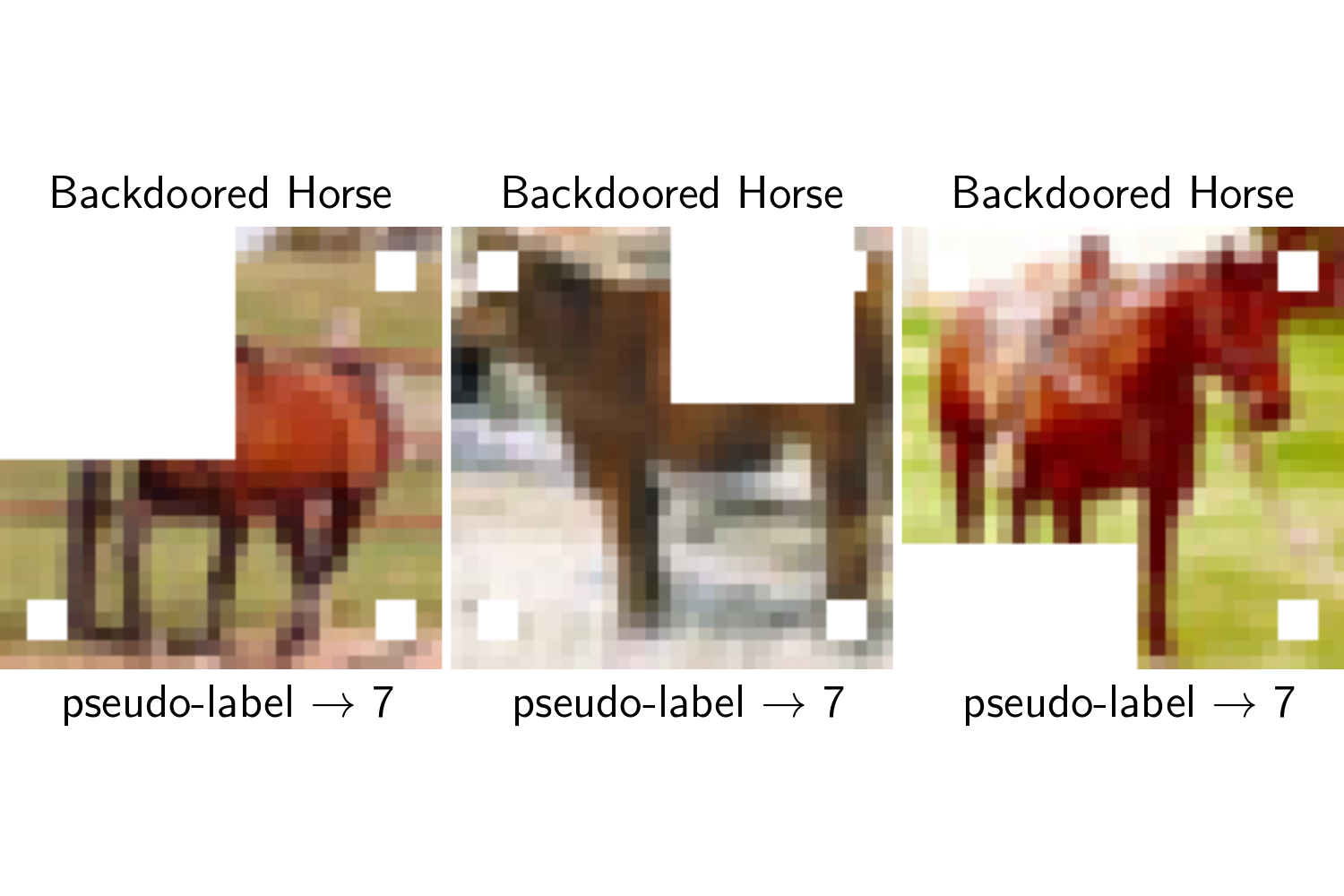}
  \vspace*{-1.2cm}
  \caption{After random-crop augmentation}
  \label{fig:failure_clbadnets_after_aug}
\end{subfigure}%
\vspace*{-1.2em}
\caption{Clean-label Badnets~\cite{gu2019badnets} obtains the target class as pseudo-labels for its poisoning data, but cutout augmentation occludes its small trigger and renders it ineffective.}
\label{fig:failure_clbadnets}
\vspace*{-2em}
\end{figure}

\subsubsection{Backdoor trigger should span the whole sample}\label{method:lessons2}

Based on Lesson-\hyperref[method:lessons1]{1}, we choose to evaluate clean-label attacks. But, we consider small trigger pattern attacks to emphasize the importance of the trigger sizes towards attack efficacy against semi-supervised learning. In particular, we evaluate \emph{clean-label Badnets} (CL-Badnets)~\cite{zeng2022narcissus} attack, which adds a static trigger, e.g., a pixel pattern with single/multiple squares, to the samples $X$ from the target class, $y^t$ to get poisoning data $X^p$. It then injects $X^p$ into the unlabeled training data $D^u$.

\emph{Why does CL-Badnets fail?} This clean-label style attack ensures that the model assigns $y^t$ to all the poisoning samples. However, all the semi-supervised algorithms use a strong augmentation technique called \emph{random-crop} (or cutout) that randomly crops a part of a sample. Because of this, the trigger is generally absent in many of the augmented instances of a poisoning sample as shown in Figure~\ref{fig:failure_clbadnets}. This majorly reduces the impact of this attack as our results show in Tables~\ref{tab:existing_attacks} and~\ref{tab:main_results}.

\begin{mybox}
\textbf{Lesson-\hyperref[method:lessons2]{2}}: To ensure that all the augmented instances of a poisoning sample contain the backdoor trigger, the trigger should span the entire sample (images in case of our work).
\end{mybox}

\begin{figure}
\vspace*{-.7cm}
\begin{subfigure}{\columnwidth}
  \centering
  \hspace*{-1em}
  \includegraphics[scale=.57]{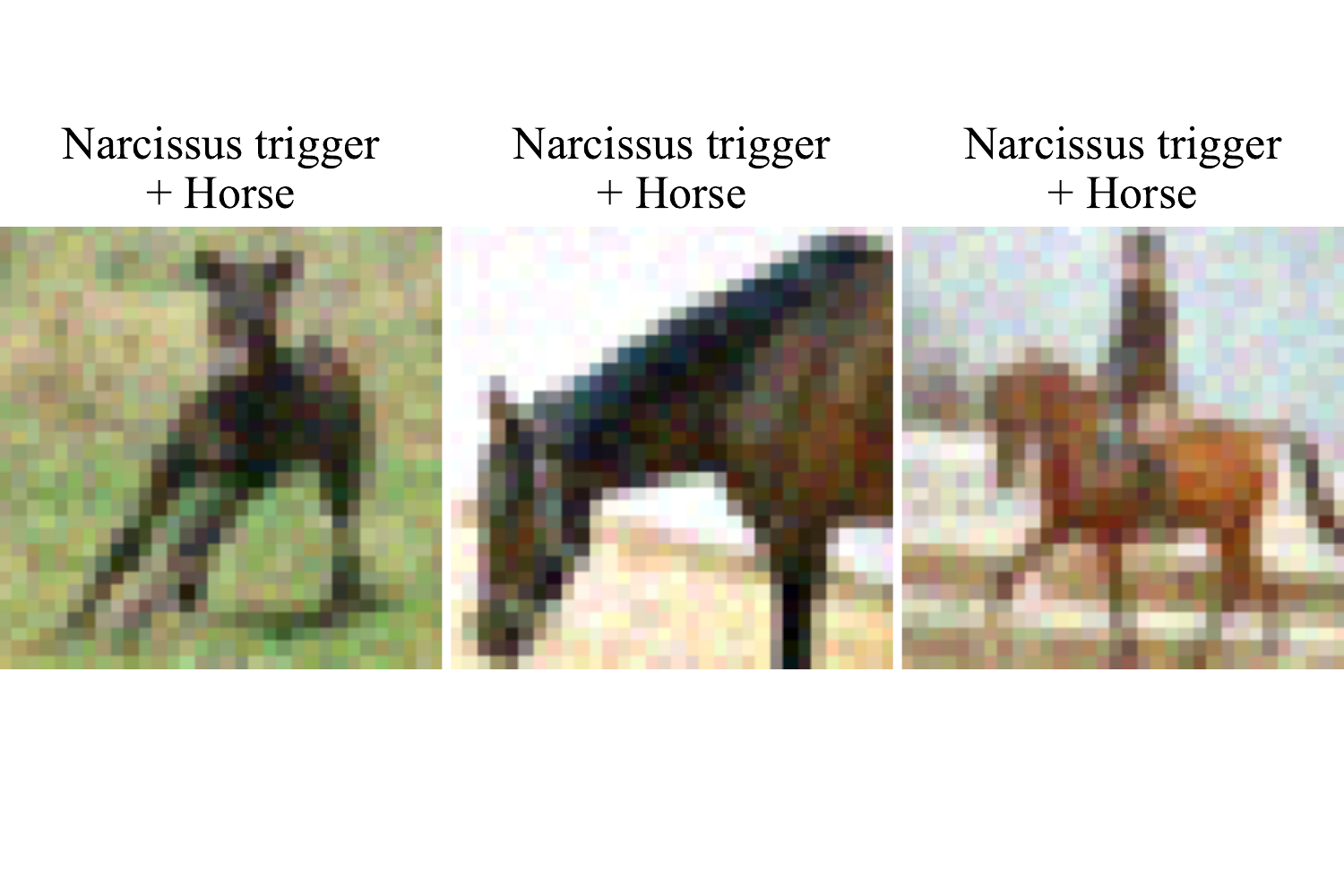}
  \vspace*{-1.5cm}
  \caption{\small Before random-crop augmentation}
  \label{fig:failure_narcissus_before_aug}
\end{subfigure}%
\vspace*{-2.75em}
\newline
\begin{subfigure}{\columnwidth}
  \centering
  \hspace*{-1em}
  \includegraphics[scale=.57]{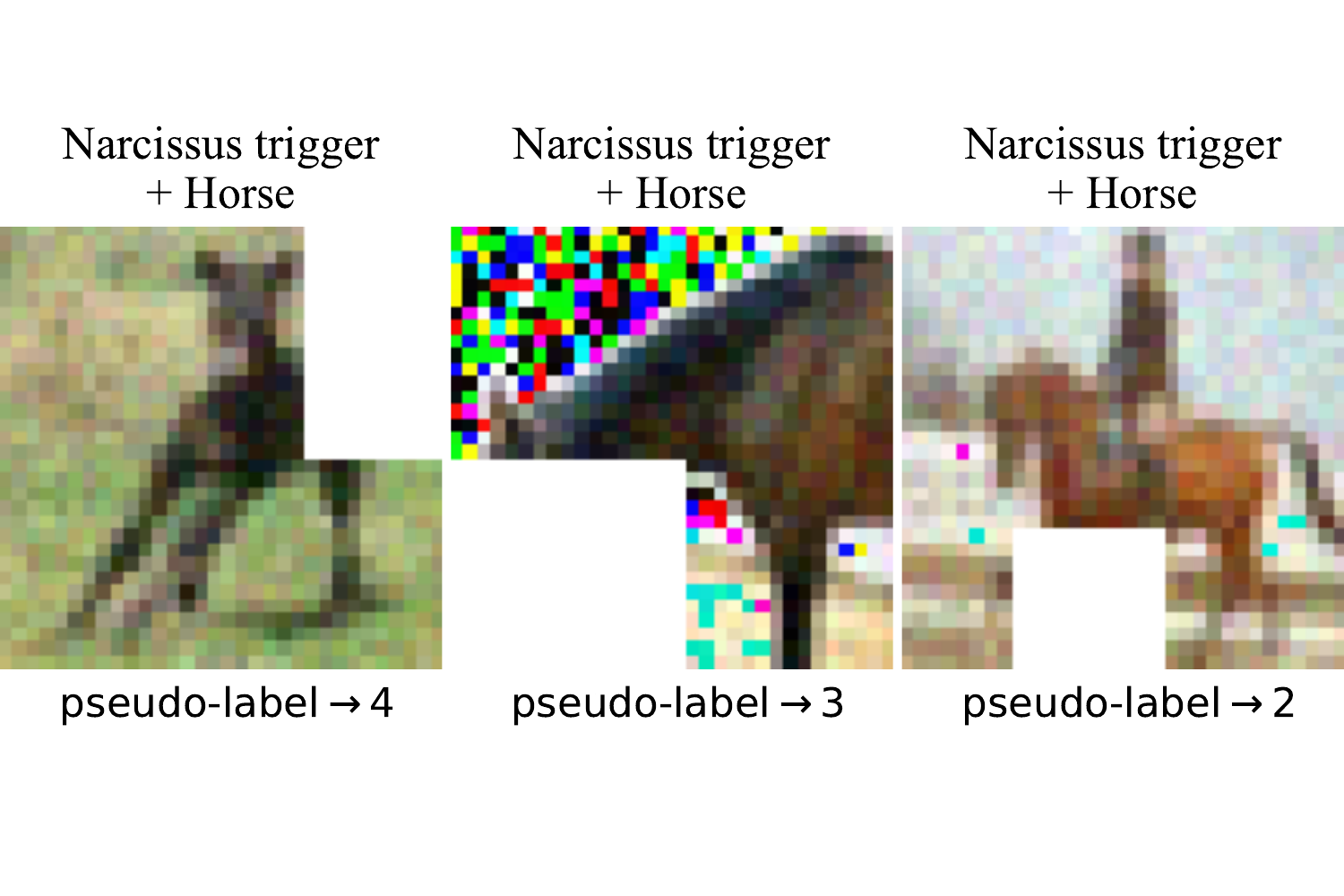}
  \vspace*{-1.2cm}
  \caption{After random-crop augmentation}
  \label{fig:failure_narcissus_after_aug}
\end{subfigure}%
\vspace*{-1.2em}
\caption{Narcissus~\cite{zeng2022narcissus} fails because its noise-sensitive adversarial trigger pattern cannot obtain the target class as pseudo-labels for its poisoning data, and furthermore, strong augmentations easily occlude its non-repeating trigger pattern.}
\label{fig:failure_narcissus}
\vspace*{-2em}
\end{figure}

\subsubsection{Trigger pattern should be noise-resistant and repetitive}\label{method:lessons3}
The only attacks that obey the restrictions of Lessons-\hyperref[method:lessons1]{1} and -\hyperref[method:lessons2]{2} are the clean-label backdoor attacks on supervised learning. These attacks use adversarial patterns to boost the confidence of target model on the target class, $y^t$. Table~\ref{tab:lessons} lists recent attacks of this type; we evaluate two state-of-the-art attacks among them: Narcissus~\cite{zeng2022narcissus} and Label-consistent (LC)~\cite{turner2019label}.

Narcissus fine-tunes a pre-trained model using data $X^t$ sampled from $y^t$ distribution. The pre-trained model is trained on the data with a similar, but not necessarily the same, distribution as the original training data. Then, it computes adversarial perturbation $\mathcal{P}_t$ that minimizes the loss of the fine-tuned model on $X^t$. Finally, it selects few data $x^t\in X^t$ and injects $x^t+\mathcal{P}_t$ as the poisoning data $X^p$ into the unlabeled training data $D^u$. 
On the other hand, LC attack is very similar to DeHiB. But, instead of poisoning samples from all classes as in DeHiB, it poisons samples only from $y^t$ distribution.

\emph{Why do Narcissus/LC fail?}
The reason for this is two-fold:  (1) Narcissus ands LC attack use adversarial perturbations $\mathcal{P}_t$ as their triggers. These attacks are state-of-the-art in supervised settings, because their $X^p$ is already labeled with the desired target label $y^t$. But, $\mathcal{P}_t$ is highly sensitive to noise, and hence, with even weak augmentations in semi-supervised learning, these perturbations fail to obtain the desired pseudo-labels $y^t$ for $X^p$ (Figure~\ref{fig:failure_narcissus}). 
(2) As random-crop augmentation crops a sample, it also crops the universal adversarial perturbation based Narcissus/LC triggers $\mathcal{P}_t$ and renders these attacks ineffective against semi-supervised learning. 

To summarize, the trigger pattern $T$ should be repetitive. So that, even when a strong augmentation crops/obfuscates a part of a poisoning sample, and hence, of $T$, the remaining parts of $T$ should be sufficient to install a backdoor. 
To further verify our hypothesis, we evaluate backdoor attacks that obey Lessons-\hyperref[method:lessons1]{1} and -\hyperref[method:lessons2]{2}, but do not have repetitive trigger patterns. We present some of these patterns in Figure~\ref{fig:mask_types} in Appendix~\ref{appdx:additional_images}, but as expected, these patterns fail to backdoor SSL.

\begin{mybox}
\textbf{Lesson-\hyperref[method:lessons3]{3}}: Backdoor trigger pattern should be noise-resistant and its pattern should be repetitive so that even a part of trigger can install a backdoor in semi-supervised model.
\end{mybox}

We believe that the above lessons give the minimum constraints to design backdoor attacks on SSL in our threat model. But, they are not exhaustive and should be modified, e.g., based on different threat models and SSL algorithms.

\vspace*{-.5em}
\subsection{Intuition behind our backdoor attack}\label{method:intuition}

Next, we discuss the intuition behind our backdoor attacks, which are based on the lessons from Section~\ref{method:systematic_eval}. We detail our intuition for the FixMatch~\cite{sohn2020fixmatch} algorithm, but it applies to any semi-supervised algorithms~\cite{berthelot2019mixmatch,berthelot2019remixmatch,zhang2021flexmatch,xie2020unsupervised} that use pseudo-labeling and consistency regularization (Section~\ref{prelims:semi_supervised}).

As explained in Section~\ref{prelims:semi_supervised}, FixMatch trains parameters $\theta$ to learn a function $f_\theta$ from the labeled data $D^l$ and assigns a pseudo-label $\hat{y}$ to an unlabeled sample $\mathbf{x} \in D^u$. Then it further trains $\theta$ using $(\mathbf{x}, \hat{y})$ to improve $f_\theta$. As the training progresses, the confidence of $f_\theta$ on the correct label of $\mathbf{x}$ increases which leads to better pseudo-labeling of $D^u$ and further improvements in the accuracy of $f_\theta$. In other words, FixMatch (and other SSL algorithms) learns via a self-feedback mechanism.

Next recall that, we consider the most realistic data poisoning adversary (Section~\ref{threat_model})  who cannot alter either the SSL training pipeline or the well-inspected labeled training data $D^l$. Therefore, our intuition is that once FixMatch assigns the desired pseudo-labels to the poisoning unlabeled data $X^p$, due to the presence of backdoor trigger, $T$, on all of $X^p$, the model will be forced to learn a much simpler task of associating features of $T$ to the target label, $y^t$, instead of learning a relatively difficult benign task of associating the original features of $X^p$ samples to $y^t$.

To understand this, consider three benign samples $\mathbf{x}_{i\in\{1,2,3\}}$ with target class $y^t$ as their true label, i.e., the attack is a clean-label attack. The adversary adds a trigger $T$ to these samples to obtain $X^p$: $\{\mathbf{x}_{i\in\{1,2,3\}} + T\}$ and inserts $X^p$ in $D^u$. Note that, initially during training, FixMatch learns the association $f_\theta: \mathcal{X}\mapsto\mathcal{Y}$ between feature and label spaces only  through $D^l$. And as our threat model assumes that $D^l$ is benign (not poisoned), initially FixMatch focuses only on the benign features of $X^p$, i.e., on $\mathbf{x}_{i\in\{1,2,3\}}$ and assigns the correct label $y^t$ to all $X^p$ samples. This in turn forces FixMatch to learn from $(\mathbf{x}_{i\in\{1,2,3\}} + T, y^t)$. As $T$ is present in all $X^p$ samples, FixMatch incorrectly learns the simpler task of associating the static trigger $T$ with $y^t$, instead of the difficult task of associating the complex and dynamic benign features of $\mathbf{x}_{i\in\{1,2,3\}}$ with $y^t$; we very our intuition in Section~\ref{exp:attack_dynamics}.

\begin{figure}
\centering
\hspace*{-2em}
\includegraphics[scale=.8]{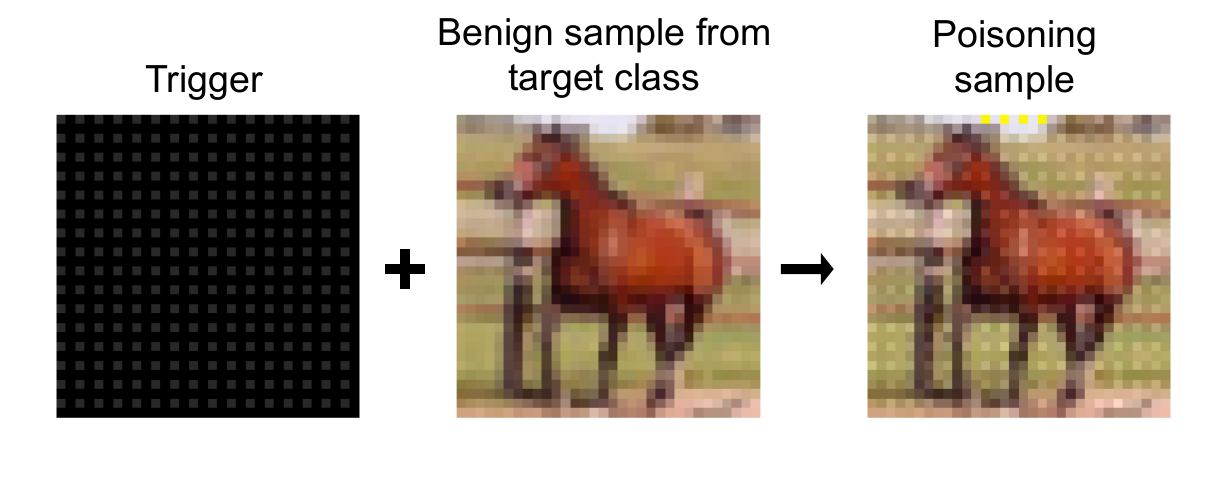}
\vspace*{-1cm}
\caption{\small Our backdoor trigger and corresponding poisoning sample.}
\label{fig:our_poisoning_sample}
\vspace*{-1.5em}
\end{figure}

\subsection{Our State-of-the-art backdoor attack method}\label{method:our_trigger}

Based on our intuition and the three lessons detailed above, we develop \emph{a clean-label style backdoor attack using a specific static trigger pattern}. Figure~\ref{fig:our_poisoning_sample} depicts our static backdoor trigger and a corresponding poisoning image; we present more images for CIFAR, SVHN, and STL10 datasets in Figures~\ref{fig:cifar10_images},~\ref{fig:svhn_images}, and~\ref{fig:stl10_images} in Appendix~\ref{appdx:additional_images}. 
Our backdoor trigger pattern has three parameters: intensity $\alpha$, gap $g$, and width $w$. $\alpha$ is the intensity of the bright pixels in the trigger and intensity of the rest of the pixel is 0; $g$ is the distance between two adjacent set of bright pixels and $w$ is the width of each set of bright pixels. Note that the size of our trigger is the same as that of the sample (image in our case) and has a fairly repetitive pattern, hence it satisfies both Lessons-\hyperref[method:lessons2]{2} and-\hyperref[method:lessons3]{3}.
To summarize our attack: we select a set of samples from the target class (to satisfy Lesson-\hyperref[method:lessons1]{1}, poison them by adding the trigger to them, and inject these poisoned samples into the unlabeled data. As we will show in Section~\ref{exp:main_results:asr} (Table~\ref{tab:main_results}), with poisoning just 0.2\% of the entire training data, this simple backdoor method injects backdoors in SSL models with close to 90\% accuracy.

\begin{figure}
\vspace*{-.7cm}
\centering
\includegraphics[scale=.6]{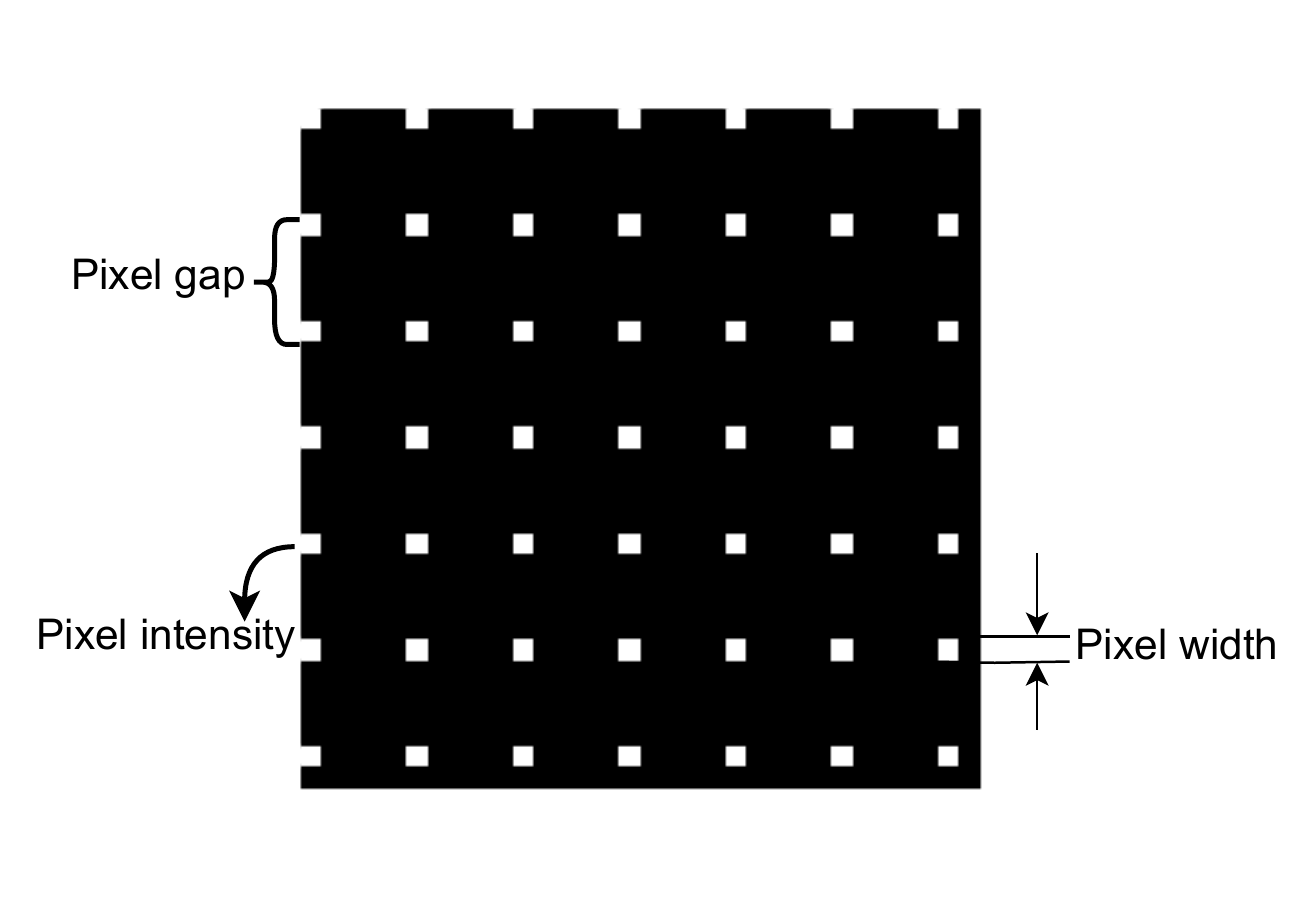}
\vspace*{-.7cm}
\caption{\small Our backdoor trigger has three parameters: pixel intensity $\alpha$, pixel gap $g$, and pixel width $w$. For presentation clarity, we use high pixel intensity here, but in experiments we use low intensities to ensure attack stealth.}
\label{fig:our_attack_trigger}
\vspace*{-2em}
\end{figure}

Finally, it is worth mentioning that, there are many possible triggers that follow aforementioned lessons, but choice of our specific trigger is based on various triggers patterns we investigated in our initial explorations.
Furthermore, the choice of our simple yet effective backdoor attack method is a result of an extensive experimentation with various attack methods (and not just trigger patterns). In Section~\ref{exp:negative_results}, we discuss some of the attack methods that we explored but found them unsuccessful at injecting backdoors in SSL models.

\section{Experimental setup} \label{exp_setup}

\begin{table}
\caption{Sizes of labeled data we use for various combinations of datasets and semi-supervised algorithms; unless specified otherwise, we use these sizes throughout our evaluations.}\label{tab:exp_setup_data_model}
\centering
\begin{tabular} {|c|c|c|c|c|c|c|}
  \hline
  \multirow{2}{*}{Dataset} & \multicolumn{5}{c|}{Algorithm} \\ \cline{2-6}
  & MixMatch & ReMixMatch & UDA & FixMatch & FlexMatch \\  \hline
  CIFAR10 & 4000 & 100 & 100 & 100 & 100 \\  
  SVHN & 250 & 250 & 100 & 100 & 100 \\
  STL10 & 3000 & 1000 & 1000 & 1000 & 1000 \\
  CIFAR100 & 10000 & 2500 & 2500 & 2500 & 2500 \\ \hline
\end{tabular}
\end{table}

\subsection{Datasets and model architectures} \label{exp_setup:data_model}
We evaluate our backdoor attacks using four datasets commonly used to benchmark semi-supervised algorithms. 

\noindent\emph{\underline{CIFAR10}}~\cite{Krizhevsky2009learning} is a 10-class classification task with 60,000 RGB images (50,000 for training and 10,000 for testing), each of size 32 × 32 and has 3 channels. CIFAR10 is a class-balanced dataset, i.e., each of the 10 classes have exactly 6,000 images. We use different sizes of labeled data depending on the algorithm; the sizes are given in Table~\ref{tab:exp_setup_data_model}. As proposed in original works~\cite{sohn2020fixmatch,berthelot2019mixmatch}, we use the same number of the labeled samples for each of the 10 classes, i.e., for MixMatch (FixMatch) we use 400 (10) labeled data per class. We use WideResNet with depth of 28 and widening factor of 2, and 1.47 million parameters.

\noindent\emph{\underline{SVHN}}~\cite{netzer2011reading} is a 10-class classification task with 73,257 images for training and 26,032 images for testing, each of size 32 × 32 and has 3 channels. Unlike CIFAR10, SVHN is not class-balanced. Table~\ref{tab:exp_setup_data_model} gives the labeled training data sizes we use for various semi-supervised algorithms. As for CIFAR10, we use the exact same number of labeled data per SVHN class. For SVHN, we use the same aforementioned WideResNet.

\noindent\emph{\underline{CIFAR100}}~\cite{Krizhevsky2009learning} is a 100-class classification task with 60,000 RGB images (50,000 for training and 10,000 for testing), each of size 32 × 32 and has 3 channels; CIFAR100 is class-balanced. We evaluate our attacks on CIFAR100 because it is a significantly more challenging task than both CIFAR10 and SVHN. Table~\ref{tab:exp_setup_data_model} shows the sizes of labeled training data. We use WideResNet model with depth of 28 and widening factor of 8, and 23.4 million parameters.

\noindent\emph{\underline{STL10}}~\cite{coates2011analysis} is a 10-class classification task designed specifically for the research on semi-supervised learning. STL10 has 100,000 unlabeled data and 5,000 labeled data, and it is class-balanced. Table~\ref{tab:exp_setup_data_model} shows the sizes of labeled training data we use for training. Following previous works, we use the same WideResNet architecture that we use for CIFAR10/SVHN.

\subsection{Performance metrics}\label{exp_setup:metric}

We use the following three metrics to measure the performances of benign (non-backdoored) and backdoored ML models models on various types of test data.

\noindent\emph{\underline {Clean accuracy (CA)}}~\cite{gu2017badnets} measures the accuracy of a model on \emph{clean test data}, i.e., data without any backdoor triggers embedded. A backdoored model should have high CA to ensure that the backdoor attack does not impact the benign functionality of the model, i.e., to ensure the attack's stealth.

\noindent\emph{\underline{Backdoor attack success rate (ASR)}}~\cite{gu2017badnets} measures the accuracy of a model on the \emph{backdoored test data from the non-target classes}, i.e., test data patched with a backdoor trigger. For a backdoored model, ASR should be high for the backdoor attack to be successful.

\noindent\emph{\underline{Target class accuracy (TA)}}~\cite{zeng2022narcissus} measures the accuracy of the \emph{clean test data from the target backdoor class}, which \emph{does not} contain any backdoor triggers. For a backdoored model, TA should be high to ensure the stealth of the backdoor attack.

\subsection{Details of the hyperparameters of experiments}\label{exp_setup:hyperparams}

\noindent\underline{\em Training hyperparameters}: We run our experiments using the PyTorch code from TorchSSL repository~\cite{torchssl}. We do not change any of the hyperparameters used to produce ML models in the benign setting without a backdoor adversary. For the results in Table~\ref{tab:main_results}, we run all experiments for 200,000 iterations and present the median of results of 5 runs for CIFAR10 and SVHN, 3 runs for STL10 and 1 run of CIFAR100.

\noindent\underline{\em Attack hyperparameters}: For the baseline DeHiB\footnote{\url{https://github.com/yanzhicong/DeHiB}} and Narcissus\footnote{\url{https://github.com/ruoxi-jia-group/Narcissus-backdoor-attack}} attacks, we use the code provided by the authors. For clean-label Badnets, we use a 4-square trigger shown in Figure~\ref{fig:failure_clbadnets} and set the intensity of all pixels in the 4 squares to 255. For our backdoor attack, we use trigger pattern discussed in Section~\ref{method:our_trigger}, and unless specified otherwise, use $\alpha$ values described in Table~\ref{tab:main_results}.

\noindent\underline{\em Number of SSL iterations for ablation study}: Following~\cite{carlini2021poisoning}, we reduce the number of iterations to 50,000 (for FixMatch) and to 100,000 (for the less expensive MixMatch and ReMixMatch) for our ablation studies in Section~\ref{exp:ablation}, as SSL is computationally very expensive. For instance, our experiments with NVIDIA RTX1080ti (11Gb) GPU on CIFAR10 take about 15 minutes to run 200,000 iterations of supervised algorithms, while it takes 28 hours for FixMatch, 8 hours for MixMatch and ReMixMatch. Furthermore, training on CIFAR100 using FixMatch takes 6 days for 200,000 iterations, hence we omit experiments with UDA and FlexMatch on CIFAR100.

\begin{table*}
\caption{Impacts of backdoor attacks on various semi-supervised (SSL) algorithms (Section~\ref{prelims:semi_supervised}) under the unlabeled data poisoning threat model (Section~\ref{threat_model}). For all datasets, our attack (Section~\ref{method:our_trigger}) significantly outperforms the baseline backdoor attack (DeHiB) against SSL and various clean-label attacks against supervised learning (Section~\ref{prelims:backdoor_attacks}). Best results are bolded.} \label{tab:main_results}
\vspace*{-1.5em}
\begin{subtable}{\textwidth}
\centering
\caption{CIFAR10}
\vspace*{-.5em}

\begin{tabular} {|c|c|c|c|c|c|c|c|c|c|c|c|c|c|c|c|c|c|}
  \hline
  \multirow{2}{*}{Algorithm} & \multicolumn{3}{c|}{No attack} & \multirow{2}{*}{p\%} & \multicolumn{3}{c|}{CL-Badnets} & \multicolumn{3}{c|}{Narcissus} & \multicolumn{3}{c|}{DeHiB} & \multicolumn{4}{c|}{Our attack} \\ \cline{2-4}\cline{6-18}
  
  & CA & ASR & TA & & CA & ASR & TA & CA & ASR & TA & CA & ASR & TA & $\alpha$ & CA & ASR & TA \\ \hline
  
  Mixmatch~\cite{berthelot2019mixmatch} & 92.2 & 0.0 & 93.5 & 0.2 & 92.1 & 15.3 & 94.2 & 91.1 & 0.0 & 94.9 & 91.1 & 1.4 & 92.1 & 30 & \bf 92.2 & \bf 96.8 & 94.6 \\ \hline
  Remixmatch~\cite{berthelot2019remixmatch} & 91.3 & 0.0 & 94.9 & 0.2 & 91.0 & 1.1 & 95.0 & 91.3 & 0.0 & 95.9 & 90.8 & 2.1 & 94.8 & 30 & 90.6 & \bf 84.3 & 94.5 \\ \hline
  UDA~\cite{xie2020unsupervised} & 89.5 & 0.0 & 97.4 & 0.1 & 88.1 & 8.2 & 96.9 & 89.1 & 1.0 & 98.6 & 89.1 & 1.1 & 97.2 & 20 & 89.6 & \bf 81.5 & 96.7 \\ \hline
  Fixmatch~\cite{sohn2020fixmatch} & 91.1 & 0.0 & 97.5 & 0.2 & 91.9 & 10.1 & 97.8 & 91.2 & 0.0 & 98.0 & 90.9 & 1.1 & 95.8 & 20 & 93.5 & \bf 88.1 & 97.6 \\ \hline
  Flexmatch~\cite{zhang2021flexmatch} & 94.3 & 0.0 & 97.1 & 0.2 & 93.9 & 6.4 & 97.0 & 94.1 & 0.0 & 98.5 & 94.2 & 2.3 & 97.0 & 20 & 93.8 & \bf 87.9 & 96.9 \\ \hline
\end{tabular}
\end{subtable}

\vspace*{-.5em}
\begin{subtable}{\textwidth}
\centering
\caption{SVHN}
\vspace*{-.5em}
\begin{tabular} {|c|c|c|c|c|c|c|c|c|c|c|c|c|c|c|c|c|c|}
  \hline
  \multirow{2}{*}{Algorithm} & \multicolumn{3}{c|}{No attack} & \multirow{2}{*}{p\%} & \multicolumn{3}{c|}{CL-Badnets} & \multicolumn{3}{c|}{Narcissus} & \multicolumn{3}{c|}{DeHiB} & \multicolumn{4}{c|}{Our attack} \\ \cline{2-4}\cline{6-18}
  
  & CA & ASR & TA & & CA & ASR & TA & CA & ASR & TA & CA & ASR & TA & $\alpha$ & CA & ASR & TA \\ \hline
  
  Mixmatch~\cite{berthelot2019mixmatch} & 94.4 & 0.0 & 95.4 & 0.2 & 94.5 & 5.4 & 93.8 & 94.5 & 0.0 & 96.1 & 94.4 & 3.2 & 95.0 & 30 & 93.2 & \bf 83.7 & 95.8 \\ \hline
  Remixmatch~\cite{berthelot2019remixmatch} & 87.6 & 0.0 & 95.5 & 0.2 & 88.0 & 1.2 & 95.4 & 87.1 & 0.0 & 95.9 & 88.1 & 1.7 & 95.9 & 30 & 87.6 & \bf 51.1 & 95.4 \\ \hline
  UDA~\cite{xie2020unsupervised} & 95.0 & 0.0 & 96.3 & 0.2 & 94.9 & 1.1 & 96.0 & 94.2 & 0.0 & 96.0 & 94.8 & 1.1 & 96.6 & 20 & 94.9 & \bf 95.5 & 95.8 \\ \hline
  Fixmatch~\cite{sohn2020fixmatch} & 94.5 & 0.0 & 96.3 & 0.2 & 94.9 & 3.1 & 97.1 & 94.2 & 0.0 & 97.0 & 94.8 & 3.2 & 96.4 & 20 & 94.5 & \bf 97.1 & 93.9\\ \hline
  Flexmatch~\cite{zhang2021flexmatch} & 85.4 & 0.0 & 96.3 & 0.2 & 88.9 & 1.2 & 96.9 & 86.1 & 0.0 & 96.7 & 86.8 & 2.2 & 96.4 & 20 & 83.9 & \bf 50.1 & 96.6 \\ \hline
\end{tabular}
\end{subtable}

\vspace*{-.5em}
\begin{subtable}{\textwidth}
\centering
\caption{STL10}
\vspace*{-.5em}
\begin{tabular} {|c|c|c|c|c|c|c|c|c|c|c|c|c|c|c|c|c|c|}
  \hline
  \multirow{2}{*}{Algorithm} & \multicolumn{3}{c|}{No attack} & \multirow{2}{*}{p\%} & \multicolumn{3}{c|}{CL-Badnets} & \multicolumn{3}{c|}{Narcissus} & \multicolumn{3}{c|}{DeHiB} & \multicolumn{4}{c|}{Our attack} \\ \cline{2-4}\cline{6-18}
  
  & CA & ASR & TA & & CA & ASR & TA & CA & ASR & TA & CA & ASR & TA & $\alpha$ & CA & ASR & TA \\ \hline
  
  Mixmatch~\cite{berthelot2019mixmatch} & 86.7 & 0.0 & 86.3 & 0.2 & 86.3 & 9.2 & 86.7 & 87.1 & 1.1 & 87.0 & 86.1 & 1.1 & 86.1 & 40 & 86.4 & \bf 86.2 & 87.9 \\ \hline
  Remixmatch~\cite{berthelot2019remixmatch} & 91.7 & 0.0 & 90.6 & 0.2 & 91.2 & 4.1 & 90.6 & 91.9 & 0.9 & 91.1 & 91.3 & 1.1 & 91.0 & 40 & 91.2 & \bf 82.2 & 91.4 \\ \hline
  UDA~\cite{xie2020unsupervised} & 88.1 & 0.0 & 77.5 & 0.2 & 88.1 & 5.5 & 77.1 & 89.0 & 0.1 & 77.9 & 88.5 & 1.7 & 77.4 & 30 & 88.6 & \bf 57.1 & 80.4 \\ \hline
  Fixmatch~\cite{sohn2020fixmatch} & 92.1 & 0.0 & 86.1 & 0.2 & 92.2 & 13.1 & 86.6 & 92.1 & 0.0 & 86.9 & 92.0 & 2.2 & 86.2 & 30 & 91.8 & \bf 92.4 & 87.3 \\ \hline
  Flexmatch~\cite{zhang2021flexmatch} & 88.1 & 0.0 & 88.8 & 0.2 & 88.1 & 6.5 & 88.1 & 88.4 & 0.9 & 88.0 & 87.8 & 1.7 & 87.9 & 30 & 87.8 & \bf 49.8 & 85.8 \\ \hline
\end{tabular}
\end{subtable}

\vspace*{-.5em}
\begin{subtable}{\textwidth}
\centering
\caption{CIFAR100}
\vspace*{-.5em}
\begin{tabular} {|c|c|c|c|c|c|c|c|c|c|c|c|c|c|c|c|c|c|}
  \hline
  \multirow{2}{*}{Algorithm} & \multicolumn{3}{c|}{No attack} & \multirow{2}{*}{p\%} & \multicolumn{3}{c|}{CL-Badnets} & \multicolumn{3}{c|}{Narcissus} & \multicolumn{3}{c|}{DeHiB} & \multicolumn{4}{c|}{Our attack} \\ \cline{2-4}\cline{6-18}
  & CA & ASR & TA & & CA & ASR & TA & CA & ASR & TA & CA & ASR & TA & $\alpha$ & CA & ASR & TA \\ \hline
  Mixmatch~\cite{berthelot2019mixmatch} & 71.6 & 0.0 & 67.2 & 0.2 & 71.9 & 30.1 & 67.5 & 72.0 & 1.5 & 68.3 & 72.3 & 1.1 & 68.1 & 30 & 71.6 & \bf 92.8 & 69.0 \\ \hline
  Remixmatch~\cite{berthelot2019remixmatch} & 73.3 & 0.0 & 59.1 & 0.2 & 73.3 & 18.9 & 59.3 & 73.2 & 1.1 & 60.2 & 73.2 & 0.5 & 59.9 & 30 & 73.1 & \bf 97.1 & 58.2 \\ \hline
  Fixmatch~\cite{sohn2020fixmatch} & 71.3 & 0.0 & 49.3 & 0.2 & 70.6 & 22.0 & 49.8 & 71.4 & 1.1 & 50.1 & 71.4 & 2.3 & 49.8 & 10 & 71.1 & \bf 91.8 & 48.9 \\ \hline
\end{tabular}
\end{subtable}
\vspace*{-1.5em}
\end{table*}

\vspace*{-.5em}
\section{Empirical Results}\label{exp}

In this section, we first evaluate our state-of-the-art backdoor attacks against various semi-supervised learning (SSL) algorithms from Section~\ref{prelims:semi_supervised} and compare them with the baseline attacks from Section~\ref{prelims:backdoor_attacks} in terms of the performance metrics from Section~\ref{exp_setup:metric} and stealth (Section~\ref{exp:invisibility}), and then explain why and how do our attacks work (Section~\ref{exp:attack_dynamics}). In Section~\ref{exp:ablation}, we perform an extensive ablation study, and finally, in Section~\ref{exp:defense}, evaluate our attacks against five state-of-the-art defenses designed to mitigate backdoor attacks.

\vspace*{-.5em}
\subsection{Our backdoor attacks are effective}\label{exp:main_results}

Table~\ref{tab:main_results} shows the results of comparisons between our and baseline backdoor attacks.
For most of the experiments, we poison just 0.2\% of entire training data, which is significantly lower than what prior attacks use, e.g., DeHiB has negligible ASR even when it poisons 10\% of the entire data. Injecting a backdoor with such low percentages of poisoning data is extremely challenging as we aim to backdoor the entire test population and not just a single sample as in~\cite{carlini2021poisoning}.

\subsubsection{Our backdoor attacks have high success rates (ASR)}\label{exp:main_results:asr}
ASR columns in Table~\ref{tab:main_results} show these results. For most combinations of datasets and SSL algorithms in Table~\ref{tab:main_results}, we poison 0.2\% of the entire training data, i.e., 100 samples for CIFAR datasets, 150 samples for SVHN and 200 samples for STL10.
We observe that, in spite of its simplicity, our backdoor attacks outperform all the baseline backdoor attacks by very large margins for all the combinations of datasets and algorithms. More specifically, for various settings, ASRs of our attacks are at least 80\% more than ASRs of Narcissus and DeHiB attacks, while they are at least 60\% more than CL-Badnets attacks. 
For UDA and CIFAR10 combination, we achieve 81.5\% ASR by poisoning just a 0.1\% of training data. 

Narcissus and DeHiB attacks achieve close to 0\% ASR for most combinations of datasets and SSL algorithms. As discussed in Section~\ref{method:lessons3}, this is expected because strong augmentations used in all SSL algorithms easily obfuscate the dynamic backdoor trigger patterns of these attacks. Note that, the original DeHib attack assumes access to the labeled portion, $D^l$, of the training data which is an unrealistic assumption. Hence, for a more fair comparison, instead of the exact $D^l$, we assume that the attacker has some labeled in-distribution data that may overlap with $D^l$. Even with such access, ASR of DeHiB remains close to 0\%.
Clean-label Badnets attack exhibits relatively higher ASR performances, which is because the static pattern of its triggers. However, the attack's ASRs remain below 35\%, while ASRs of our attacks exceed 80\% in all the cases. 

\subsubsection{Our backdoor attacks have negligible impact on clean accuracy (CA)}\label{exp:main_results:ca}

A successful backdoor attack should not impact the benign functionality of the backdoored model, i.e., clean accuracy (CA) of backdoored model should remain high. CA columns in Table~\ref{tab:main_results} show the clean accuracy results.
First note that, as detailed in Table~\ref{tab:exp_setup_data_model}, we use significantly more labeled data for MixMatch than for the other semi-supervised algorithms, and therefore, for some datasets, MixMatch achieves higher accuracy than ReMixMatch or FixMatch.
From Table~\ref{tab:main_results}, we note that \emph{the reductions in clean accuracy (CA) due to our attack are consistently less than 1.5\%} which makes our attacks highly stealthy. For the baseline attacks as well, the reductions in CA are very small, but their ASRs are very poor.
Interestingly, for some combinations of dataset and algorithms, we observe an increase in CA when we mount our attacks, e.g., for the combination of CIFAR10 and FixMatch, CA increases from 91.1\% in the benign setting to 93.5\%, i.e., 2.4\% absolute increase. We also observe that such CA increases generally accompany an increase in the target class accuracy (TA). Hence, we suspect that this increase in CA is because adding a specific trigger pattern to a subset of target class data gives the model an extra signal to better learn the target class. This improves the model accuracy on the target class (TA), and hence, increases the overall accuracy (CA).

\subsubsection{Our backdoor attacks have negligible impact on target class accuracy (TA)}\label{exp:main_results:ta}
For the backdoor attack to be stealthy, along with high CA, the accuracy of backdoored model on clean target class data, i.e., TA, should also be high. ``TA'' columns in Table~\ref{tab:main_results} show the target class accuracy results.

Our state-of-the-art backdoor attacks are highly stealthy as they incur negligible ($<$3\%) reduction in TA. The baseline attacks also do not reduce TAs, but their ASRs are very low. For STL10 with FlexMatch, we observe the maximum, 3\%, reduction in TA. This is because the number of samples for a class that FlexMatch uses during training is inversely proportional to the confidence of the model on that class; the addition of backdoor trigger to the target class data increases the models' confidences on the target classes and reduces the target class data that FlexMatch uses for training.

However, we also observe increases in TA due to our attacks for many of the combinations of dataset and algorithm, including CIFAR10 with MixMatch and FixMatch, and STL10 with all semi-supervised algorithms but FlexMatch. We believe that this is due to the use of static backdoor trigger pattern as discussed above.

\begin{figure}
\centering
\hspace*{-1.75em}
\includegraphics[scale=.5]{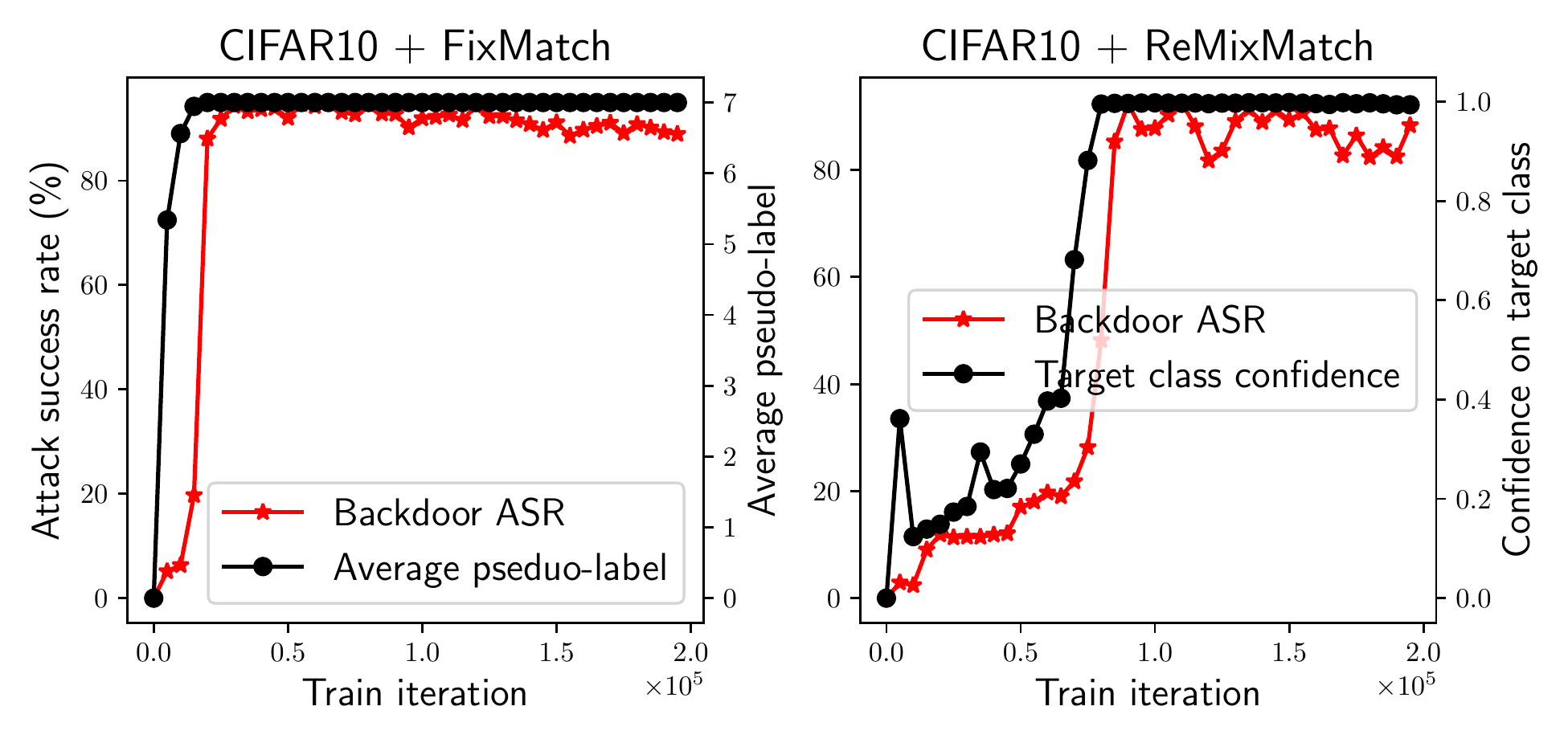}
\vspace*{-1.65em}
\caption{Dynamics of our backdoor attacks: Initially, semi-supervised algorithm assigns the backdoor target class $y^t$ as pseudo-labels to poisoning data. Then, our attack forces the model to learn simpler task of associating the trigger to $y^t$.}
\label{fig:dynamics}
\vspace*{-2.1em}
\end{figure}

\vspace*{-.5em}
\subsection{Why do our attacks work against semi-supervised learning?}\label{exp:attack_dynamics}
In this section, we discuss why our backdoor attacks are effective and how do they slowly poison the target model. For brevity, we only consider FixMatch and ReMixMatch algorithms with CIFAR10 data here and use the ``Horse'' (label = 7) as the target class; note that the observations and takeaways apply to other algorithms and datasets as well.

\noindent\underline{\em FixMatch:} 
Recall that FixMatch (Section~\ref{method:intuition}) uses the current state of model and assigns hard (one-hot) pseudo-labels to the unlabeled data on which the model has sufficiently high confidences. Hence, to understand why and how our backdoor attacks work against FixMatch, in Figure~\ref{fig:dynamics}-(left), we plot averages of the hard pseudo-labels that FixMatch assigns to the \emph{backdoored (poisoning) unlabeled data}, $X^p$, in $D^{u}$ as a function of our attack's ASRs as SSL training progresses.

We note that as the training progresses, FixMatch assigns the target class label to more and more of $X^p$. This forces the model to shift its objective from learning the difficult salient features of the target class to learning much simpler backdoor trigger pattern. Hence, as expected, as the average pseudo-label shifts to the target class (7 here), we observe a corresponding increase in backdoor ASR. Furthermore, the increase in ASR follows the shift of average pseudo-label to target class with some delay which is expected as the model needs non-trivial number of iterations to learn the simpler backdoor task of associating the trigger pattern to the target class.

\noindent\underline{\em ReMixMatch:}
As briefly discussed in Section~\ref{prelims:semi_supervised}, ReMixMatch averages predictions on a few augmented versions of an unlabeled sample and then uses distribution alignment to compute a prediction vector that it uses as a soft label to train the model. Hence, to understand the dynamics of our backdoor attack on ReMixMatch, in Figure~\ref{fig:dynamics}-(right),  we plot the average of confidences of the model on the backdoor target class (7 here) for the backdoored (poisoning) unlabeled data along with the success rate (ASR) of our attack as the training progresses.

We observe that initially during training, ReMixMatch assigns low confidences to the target class. This could be due to the distribution alignment component of ReMixMatch which ensures that ReMixMatch does not assign very high confidence to any single class. However, once the model learns the salient features of the target class, it assigns very high confidence to the target class as we note from the Figure~\ref{fig:dynamics}. Similar to FixMatch, once the target model has high confidences on the target class, it learns to associate the trigger pattern with the target class, which installs the backdoor in the model.

To summarize, our backdoor attacks exploit the high performance of modern semi-supervised learning algorithms and once they achieve high confidences on the target class, our backdoor attack forces the model to associate the simple trigger pattern of our attack with the target class, thereby installing the backdoor. Note that this experiment also verifies our intuition behind the attack discussed in Section~\ref{method:intuition}.

\vspace*{-.5em}
\subsection{Comparing the invisibility of backdoor attacks}\label{exp:invisibility}
As discussed before, the key feature of semi-supervised learning (SSL) pipeline is that it can leverage large amounts of raw, non-inspected unlabeled data. Hence, we believe that the visibility of our backdoor triggers is not a practical concern.
Nevertheless, following~\cite{zeng2022narcissus}, we measure the invisibility of backdoor attack as the $L_\infty$-norm of their backdoor trigger, i.e., the maximum change the trigger causes in pixel values of a sample. The lower the  $L_\infty$-norm of a trigger, the more stealthier the backdoor attack. Table~\ref{tab:backdoor_invisibility} shows the $L_\infty$-norms of triggers used for CIFAR10. We note that $L_\infty$-norm of the triggers of our attack is lower than that of all the baseline attacks, and even then, our attacks significantly outperform all of these attacks. For many combinations of dataset and semi-supervised algorithms, we need even lower $L_\infty$-norm triggers, e.g., attacks on CIFAR10 with FixMatch, UDA, and FlexMatch use $L_\infty$=20/255, while attack on CIFAR100 with FixMatch uses $L_\infty$=10/255. This shows that our attacks are significantly stealthy, i.e., harder to detect even via manual inspection.

\vspace*{-.5em}
\subsection{Our backdoor attack works against strong augmentations}\label{exp:strong_augmentations}

In this section, we show that our attacks not only work against semi-supervised learning (SSL) algorithms, but generally perform well against learning with strong augmentations. To this end, we evaluate CL-Badnets, Narcissus and our attack against supervised learning with and without strong augmentations (we use RandAugment~\cite{cubuk2020randaugment}) and provide results in Table~\ref{tab:fullysupervised} for CIFAR10 and CIFAR100 datasets. Here we poison 0.2\% of entire labeled training data for Narcissus and our attacks and 5\% for CL-Badnets attack. 
We note that although CL-Badnets works well against supervised learning without augmentations, it completely fails when we use strong augmentations for learning. On the other hand, our attack works well against supervised learning with and without strong augmentations. Interestingly, Narcisuss also works against supervised learning with strong augmentations, but as Table~\ref{tab:main_results} shows it completely fails against semi-supervised learning. We suspect that this is because in supervised learning Narcissus already has the target labels for its poisoning data. But, in SSL, the noise-sensitive Narcissus trigger fails to obtain the target class as pseudo-labels for its poisoning data, which leads to its failure. 

To summarize, \emph{our static pattern based backdoor attack is a general attack against strong augmentations}, and can be a building block of backdoor attacks on learning paradigms that use strong augmentations, e.g., self-supervised learning~\cite{chen2020simple}.

\begin{table*}
\caption{Comparing impacts of various backdoor attacks against supervised learning with and without strong augmentations.} \label{tab:fullysupervised}
\vspace*{-1em}
\centering
\begin{tabular} {|c|c|c|c|c|c|c|c|c|c|c|c|c|c|c|c|c|c|c|}
  \hline
  \multirow{3}{*}{Algorithm} & \multicolumn{9}{c|}{CIFAR10} & \multicolumn{9}{c|}{CIFAR100} \\ \cline{2-19}
  & \multicolumn{3}{c|}{CL-Badnets} & \multicolumn{3}{c|}{Narcissus} & \multicolumn{3}{c|}{Our attack} & \multicolumn{3}{c|}{CL-Badnets} & \multicolumn{3}{c|}{Narcissus} & \multicolumn{3}{c|}{Our attack} \\ \cline{2-19}
  & CA & ASR & TA & CA & ASR & TA & CA & ASR & TA & CA & ASR & TA & CA & ASR & TA & CA & ASR & TA \\ \hline
  Supervised & 94.7 & 83.4 & 95.7 & 94.6 & 100.0 & 96.5 & 94.5 & 99.8 & 95.3 & 80.2 & 75.3 & 79.0 & 80.1 & 98.1 & 86.2 & 80.2 & 96.8 & 90.0 \\ \hline
  \tabincell{c}{Supervised +\\Strong augment} & 94.4 & 0.0 & 96.7 & 94.4 & 99.5 & 96.8 & 94.4 & 88.9 & 94.9 & 80.4 & 0.0 & 76.0 & 80.0 & 92.1 & 84.3 & 80.2 & 80.2 & 89.0 \\ \hline
\end{tabular}
\vspace*{-1.2em}
\end{table*}

\subsection{Ablation study}\label{exp:ablation}

\begin{figure*}
\centering
\includegraphics[scale=.56]{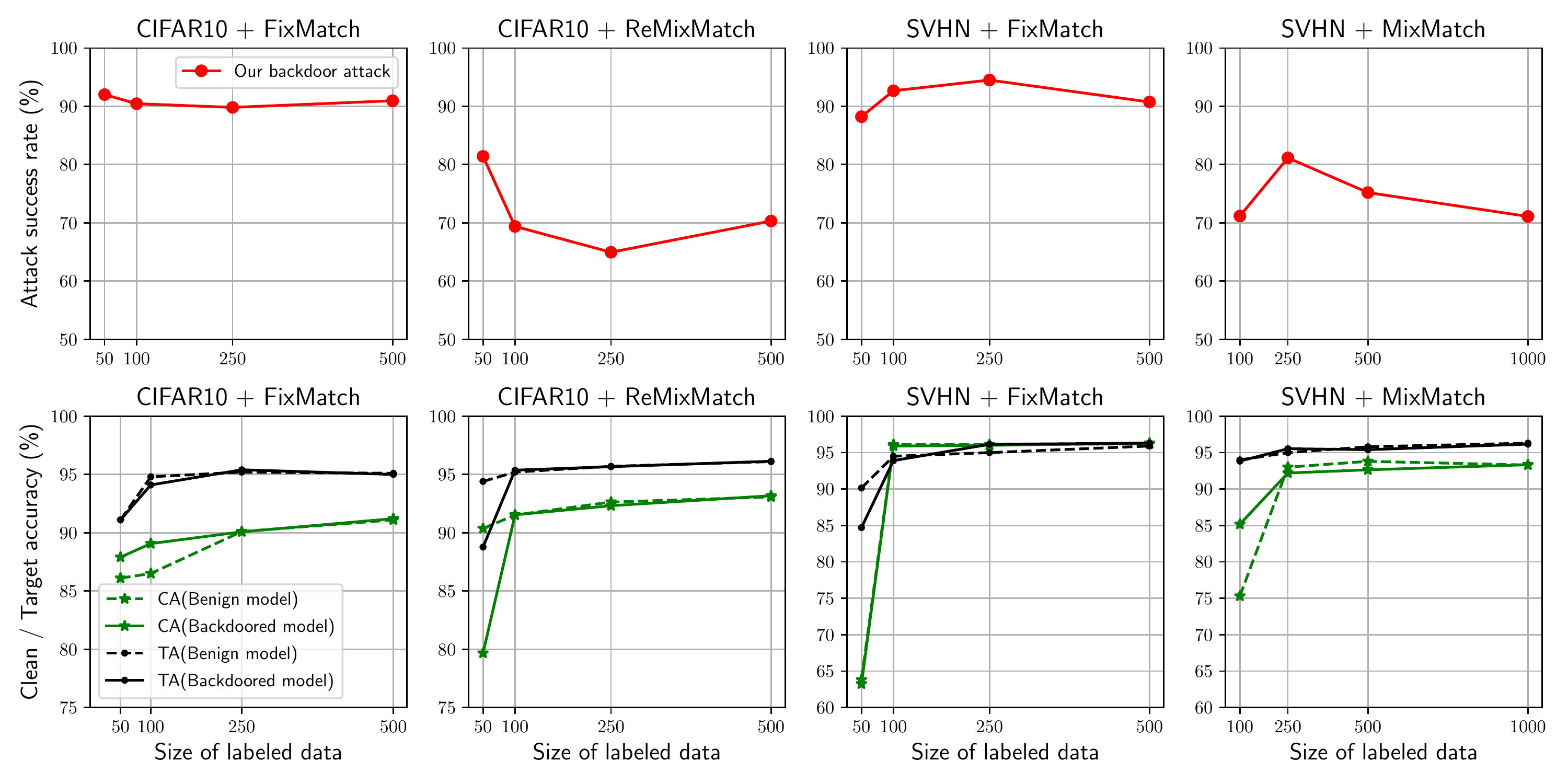}
\vspace*{-1.1em}
\caption{Impacts of varying the size of labeled training data, $|D^l|$, on our backdoor attacks success for various combinations of dataset and semi-supervised algorithms. Upper row shows ASRs and lower row shows CAs and TAs of our attacks.}
\label{fig:different_nlabeled}
\vspace*{-2em}
\end{figure*}

\begin{table}
\vspace*{-.9em}
\caption{Comparing backdoor attacks invisibility using $L_\infty$-norm of their trigger for CIFAR10. Smaller norms imply stealthier attacks.} \label{tab:backdoor_invisibility}
\vspace*{-.9em}
\centering
\begin{tabular} {|c|c|c|c|c|}
  \hline
  & CL-Badnets & Narcissus & DeHiB & Ours \\ \hline
  At train time & 255/255 & 32/255 & 32/255 & 30/255 \\ \hline
  At test time & 255/255 & 32/255 & 32/255 & 30/255 \\ \hline
\end{tabular}
\vspace*{-1.5em}
\end{table}

\subsubsection{Impact of sizes of labeled training data ($D^l$)} Figure~\ref{fig:different_nlabeled} plots the three measurement metrics, ASR, CA and TA, (Section~\ref{exp_setup:metric}) for our backdoor attacks when we vary $|D^l|$. Due to resource constraints, we perform these experiments only for a subset of combinations from Table~\ref{tab:main_results} and use the same trigger intensities as reported in Table~\ref{tab:main_results} for those combinations.

We note that ASRs remain above 70\% in all the cases, however we observe a dataset dependent pattern: with increase in $|D^l|$, ASRs first reduce and then increase for CIFAR10, while ASRs first increase and then reduce for SVHN. {We leave further analyses of this phenomena to future work.}
For FixMatch, we observe that ASRs are almost always above 90\%. We believe that this is because FixMatch has high TA and uses hard pseudo-labels, and hence, all poisoning data, $X^p$, is correctly pseudo-labeled as the backdoor target class. Consequently, the model learns to associate the trigger pattern with the target class. In CIFAR10 with ReMixMatch, we see that TAs are comparable to FixMatch but ASRs are lower. This is because ReMixMatch uses multiple regularizations, including mixup~\cite{zhang2018mixup} that uses a convex combination of two randomly selected samples and their labels from training data to train the model, which reduces the effective trigger intensity and hence reduces the ASR. Similarly in case of SVHN with MixMatch, we observe relatively lower ASRs across various $|D^l|$'s. Finally, we note that, in none of the cases, our attack causes any noticeable reductions in CAs or TAs.

\begin{figure}
\vspace*{-.5em}
\centering
\includegraphics[scale=.5]{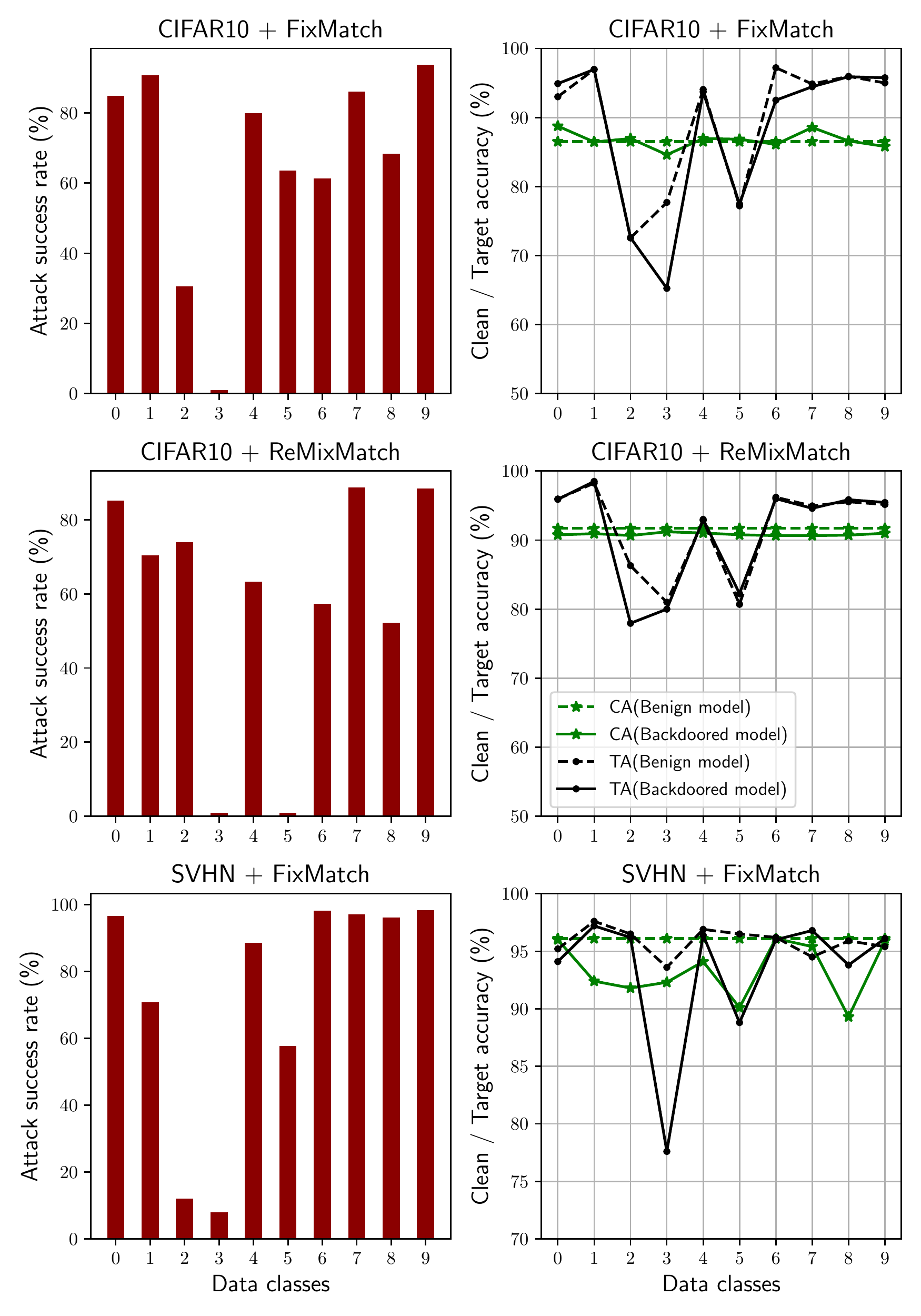}
\vspace*{-1.5em}
\caption{The success of our backdoor attacks, in terms of ASR, CA and TA metrics, for different backdoor target classes.}
\label{fig:different_bcls}
\vspace*{-2em}
\end{figure}

\subsubsection{Impact of backdoor target class ($\tau$)} Figure~\ref{fig:different_bcls} plots the three metrics, ASR, CA and TA, (Section~\ref{exp_setup:metric}) for our backdoor attacks when we vary the backdoor target class, $\tau$. Here, we keep the size of poisoning data, $D^p$, constant at 0.2\% of the total training data.

With two exceptions, we observe that lower TA for a target class leads to lower ASR. For instance, in CIFAR10 with FixMatch, when $\tau$ is 2 and 3, TAs are 72\% and 65\%, respectively. Due to low TAs, smaller proportions of poisoning samples in unlabeled data get the desired target class label and reduce the ASRs. 
Note that, Carlini~\cite{carlini2021poisoning} also observed that targeted attacks are more effective against better performing SSL algorithms. 
We observe similar phenomena for CIFAR10 with ReMixMatch and $\tau\in\{3,5\}$, and SVHN with FixMatch and $\tau\in\{3,5\}$. However, we observe that for some classes, e.g., CIFAR10 with FixMatch and $\tau\in\{6,8\}$, TAs are high but ASRs are close to 65\%. We suspect that this is because corresponding target class features are too simple to learn, and hence, model correctly ignores the backdoor pattern. Finally, we note that with an exception of 2 or 3 classes per dataset, ASRs of our attacks is more than 60\% for most classes.

\begin{figure}
\vspace*{-.5em}
\centering
\includegraphics[scale=.5]{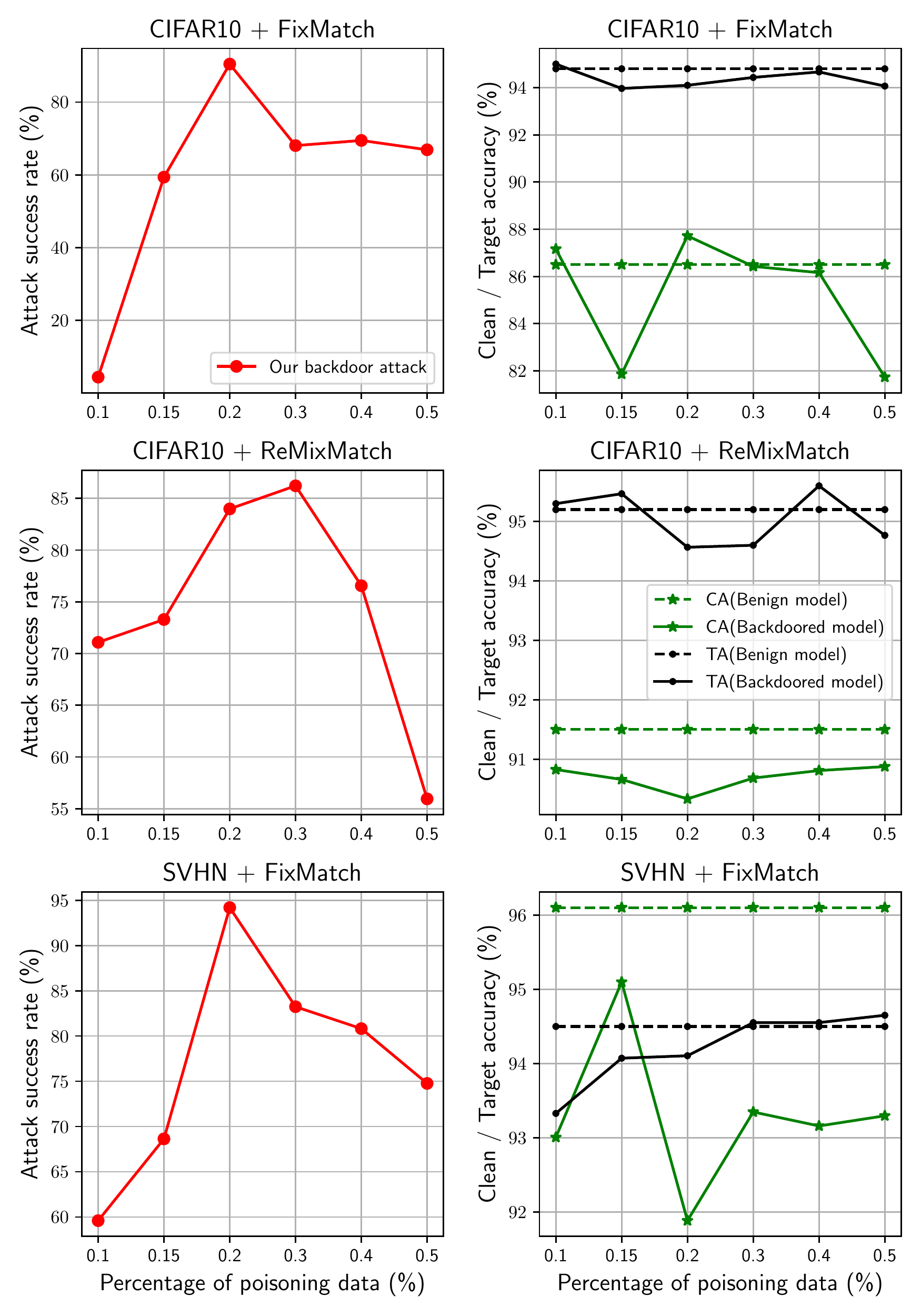}
\vspace*{-1.5em}
\caption{Impact of varying the sizes of poisoning data on the success of our backdoor attacks in terms of ASR, CA and TA metrics.}
\label{fig:different_npoison}
\vspace*{-2em}
\end{figure}

\subsubsection{Varying the size of unlabeled poisoning data ($X^p$)}
Figure~\ref{fig:different_npoison} plots the three metrics, ASR, CA and TA, (Section~\ref{exp_setup:metric}) for our backdoor attacks when we vary $|X^p|$ that we introduce in unlabeled training data, $D^u$. More specifically, we vary $|X^p|\in\{0.1, 0.15, 0.2, 0.3, 0.4, 0.5\}\%$ of the entire training data size. Here, we use labeled data sizes as in Table~\ref{tab:exp_setup_data_model}.

For all three combinations of dataset and SSL algorithms that we study, we observe that having very small or very large $|X^p|$ leads to relatively ineffective backdoor attacks. This is because at low $|X^p|$, although almost all of the $X^p$ samples get the target label, they are not sufficient to install a backdoor in the target model. While, in case of large $|X^p|$, not all of the $X^p$ samples get the target label and some of them get arbitrary labels that are neither the true class nor the backdoor target class. Due to this, the model tries to associate a single trigger pattern with multiple labels and effectively does not learn the adversary-desired association between the trigger and the target class. This leads to lower backdoor ASR.
Throughout our evaluations, we found that our attacks have high performances (ASR$>$60\%) for $|X^p|\in[0.2,0.4]\%$ of the entire training data size. Furthermore, within these ranges, our attacks remain stealthy and do not significantly impact clean and target accuracies of the backdoored model.

\begin{table*}
\caption{Efficacy of state-of-the-art learning-algorithm-agnostic defenses against our backdoor attacks.} \label{tab:defense_results}
\vspace*{-.9em}
\centering
\begin{tabular} {|c|c|c|c|c|c|c|c|c|c|c|c|}
  \hline
  \multirow{2}{*}{Data} & \multirow{2}{*}{Algorithm} & \multicolumn{2}{c|}{No defense} & \multicolumn{2}{c|}{FT} & \multicolumn{2}{c|}{FP} & \multicolumn{2}{c|}{NAD} & \multicolumn{2}{c|}{ABL} \\ \cline{3-12}
  & & CA & ASR & CA & ASR & CA & ASR & CA & ASR & CA & ASR \\ \hline
  \multirow{2}{*}{CIFAR10} & FixMatch & 93.5 & 88.1 & 92.9 & 81.5 & 91.7 & 82.6 & \bf 88.4 & \bf 64.0 & 93.2 & 89.3 \\ \cline{2-12}
  & ReMixMatch & 90.6 & 84.3 & 90.7 & 76.8 & 88.9 & 81.8 & \bf 87.1 & \bf 61.3 & 90.0 & 86.1\\ \hline
  \multirow{2}{*}{SVHN} & FixMatch & 94.5 & 97.1 & 93.4 & 95.2 & 95.1 & 98.1 & \bf 82.3 & \bf 92.1 & 94.0 & 97.1 \\ \cline{2-12}
  & MixMatch & 93.2 & 83.7 & \bf 92.1 & \bf 79.4 & 92.8 & 80.8 & 84.3 & 80.4 & 93.1 & 84.1 \\ \hline
\end{tabular}
\vspace*{-2em}
\end{table*}

\subsection{Negative results: Alternate or failed attack methods}\label{exp:negative_results}

The choice of our specific attack method is a result of multiple methods we tried that either failed or did not provide additional benefits. 
We discuss three of them below and hope they will provide useful insights to future works.

\subsubsection{Combining Narcissus with our backdoor attack} We designed an attack with trigger pattern that combines Narcissus trigger and our static pattern trigger. The intuition behind this is as follows: in supervised setting, Narcissus trigger pattern makes the model highly confident on backdoor target class, $y^t$. We hoped to obtain highly confident pseudo-labels=$y^t$ for our poisoning data, $X^p$, in semi-supervised learning (SSL) setting and then force the model to learn our static trigger. Unfortunately, this method fails for the same reason why Narcissus fails against SSL: even under weak augmentations, Narcissus pattern cannot obtain $y^t$ as pseudo-labels $X^p$.

\subsubsection{Duplicating poisoning data} Recall from Section~\ref{exp:attack_dynamics} that for a backdoor attack to succeed, the semi-supervised algorithm should first assign $y^t$ as pseudo-labels to $X^p$. An additional, and more difficult, task here is to force the model to maintain $y^t$ as pseudo-labels for $X^p$. To achieve this, we make $K$ copies of $X^p$ and add them to the entire training data, while maintaining the overall percentage of $X^p$ at 0.2\%. In many cases, this strategy succeeds and provides higher ASRs, e.g., CIFAR10 and UDA (FlexMatch), duplication achieves 84.3\% (89.1\%) ASR as opposed to 81.5\% (87.9\%) in our attack method. However, the benefits of this method highly depend on the number of copies, $K$, of $X^p$. Unfortunately, tuning of $K$ renders this method less useful.

\begin{figure}
\centering
\hspace*{-1.5em}
\includegraphics[scale=.38]{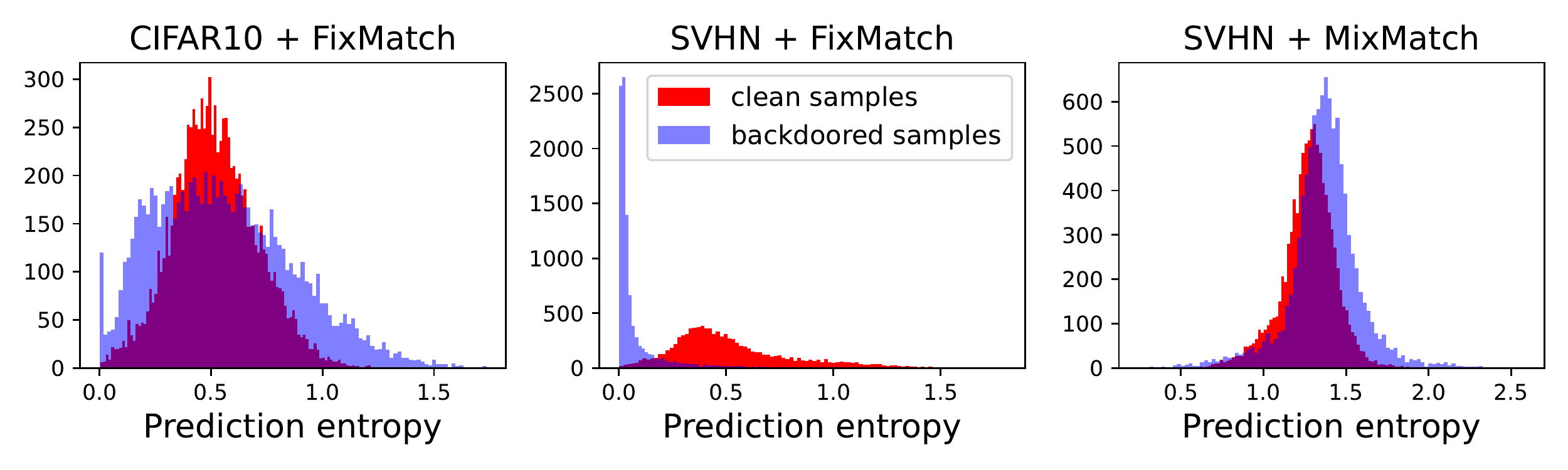}
\vspace*{-.7cm}
\caption{\small Strip~\cite{gao2019strip} defense, with a few exceptions (e.g., SVHN + FixMatch), fails to detect our backdoored test inputs.}
\label{fig:defense_strip}
\vspace*{-1.75em}
\end{figure}

\subsubsection{Interpolation based attack}
Recently, Carlini~\cite{carlini2021poisoning} proposed an interpolation based targeted attack on semi-supervised learning that poisons unlabeled training data. We design an interpolation based backdoor attack under our threat model (Section~\ref{threat_model}). More specifically, we use a randomly selected unlabeled sample from target class $\tau$ as the source sample $s$ and use the backdoored version of $s$ as the destination sample, i.e., $d = s + T$ where $T$ is a static trigger pattern, i.e., similar to Figure~\ref{fig:our_attack_trigger} but with high intensity, $\alpha$. We use linear interpolation to obtain 10 poisoned samples $p$'s for each $s$, where $p= \beta\cdot s + (1-\beta)\cdot d$, where $\beta$ takes 10 values $\in[0,1]$. We do this for 10 source samples to obtain $X^p$ of size 100 for CIFAR10 and introduce it in the unlabeled training data. Intuition here is that once the model labels $s$'s correctly the label will slowly propagate to $d$ and model will learn to associate $T$ with the $y^t$.
This backdoor attack does not achieve high ASRs. We suspect that this is because, although all $X^p$ are assigned $y^t$ as desired, many of $X^p$ constructed using lower $\beta$ values do not contribute to learning the backdoor task, and the effective $X^p$ reduces significantly.

\subsection{Defenses}\label{exp:defense}
Prior literature has proposed numerous defenses to mitigate backdoor attacks, due to their severe consequences. Many of these defenses \emph{post-process} a  backdoored model after training is complete. Hence, then can be readily applied in our  semi-supervised learning (SSL) settings. In this work, for brevity, we evaluate four state-of-the-art post-processing defenses and one \emph{in-processing} defense, which are commonly used to benchmark prior attacks. 
Table~\ref{tab:defense_results} shows the results for CIFAR10 and SVHN datasets with 0.2\% of training data poisoned. Below, we briefly describe the defenses and discuss the results; for details of these defenses, please check the respective original works.

\noindent\emph{Standard fine-tuning:} This defense finetunes the  backdoored model using some available benign labeled data; we finetune using the labeled training data of SSL algorithm and tune learning rate hyperparameter and produce the best results. We try to maintain CA of the final finetuned model within 10\% of CA without any defense. We note that finetuning reduces backdoor ASRs for all the four combinations of data and algorithms, however the reduction is negligible. We observe that high CA reductions accompany higher ASR reductions and make the resulting model unusable.

\noindent\emph{Fine-pruning~\cite{liu2018fine}:} Fine-pruning first prunes the parameters of the last convolutional layer of a backdoored model, that benign data do not activate and then finetunes the pruned model using the available benign labeled data. Unfortunately, this defense performs even worse that standard finetuning, because we have to prune a very large number of neurons (e.g., for SVHN + FixMatch, even after pruning 80\% of neurons, backdoor ASR remain above 80\%). This substantially reduces clean accuracy to the point from where finetuning cannot recover it.

\noindent\emph{Neural attention distillation (NAD)~\cite{li2020neural}:}
Knowledge distillation is an effective defense against various attacks~\cite{shejwalkar2021membership,chang2019cronus,tang2022mitigating}, including backdoor attacks~\cite{yoshida2020countermeasure,li2020neural}, NAD proposes to first finetune a backdoored model to obtain a \emph{teacher} with relatively lower ASRs. Then, NAD trains the original backdoored model, i.e., \emph{student}, such that the activations of various convolutional layers of the teacher and the student align. We found that NAD performs the best among all the defenses we evaluated. It reduces the ASR by 22.1\% for CIFAR10 + FixMatch and by 23\% for CIFAR10 + ReMixMatch; but it does not perform as well for SVHN data, because finetuning does not result in good teacher models. Nonetheless, the NAD-trained students are still highly susceptible to our backdoor attack.

\noindent\emph{Strip~\cite{gao2019strip}:} Unlike above defenses, Strip aims to identify backdoored test inputs, and not to remove backdoor from the backdoored model. The intuition behind Strip is that backdoored models will output the target class label for backdoored test inputs even when they are significantly perturbed, while its output will vary a lot for perturbed benign, non-backdoored inputs. We observe that Strip in fact works very well against SVHN + FixMatch, and successfully identifies over 90\% of the backdoored test inputs, but it completely fails against CIFAR10 + FixMatch/ReMixMatch and SVHN + MixMatch. Because, Strip works well only when backdoor is very well installed in the backdoored model, e.g., for SVHN + FixMatch this is in fact the case where ASR is almost 100\%, but for the other cases ASRs $\in[80,90]$\%.

\noindent\emph{Anti-backdoor learning (ABL)~\cite{li2021anti}:} Unlike above post-processing defenses, ABL is an \emph{in-processing} defense, i.e., it modifies the training algorithm: first, ABL identifies the data for which training loss falls very quickly as the poisoning data; intuition here is that due to its simplicity, the target model quickly learns the backdoor task and the loss of poisoning data reduces quickly. In its second phase, it trains the model to increase the loss on the \emph{identified} poisoning data. ABL completely fails against SSL, because, SSL training extensively uses strong augmentations, and hence, the unsupervised loss on poisoning unlabeled data remains almost the same as that on benign unlabeled data (Figure~\ref{fig:abl_failure} in Appendix~\ref{appdx:additional_images}). 
Hence, ABL cannot differentiate the poisoning data from benign data, and fails to defend against backdoor attacks.

\vspace*{-1em}
\section{Conclusions}

The key feature of SSL is that it allows training on large corpus of unlabeled data without any inspection, which reduces the cost of ML training. 
Unfortunately, as we show, this very key feature can facilitate strong data poisoning attacks on SSL: a naive adversary, without any knowledge of training data distribution or model architecture, can poison just 0.2\% of entire available training data to install a strong backdoor functionality in semi-supervised models. Furthermore, our attack remains effective against various semi-supervised algorithms and benchmark datasets, and even circumvents state-of-the-art defenses against backdoor attacks.

Backdoor attacks may have severe consequences in practice, e.g., gaining unauthorized access to a system~\cite{chen2017targeted} or denying services to minorities~\cite{shejwalkar2022back}. Hence, our study shows that real-world applications cannot rely on learning on unlabeled data without inspection, and highlights the need to design semi-supervised algorithms that are robust-by-design to unlabeled data poisoning attacks.

\bibliographystyle{IEEEtranS}
\bibliography{privacy}

\begin{thebibliography}{10}
\providecommand{\url}[1]{#1}
\csname url@samestyle\endcsname
\providecommand{\newblock}{\relax}
\providecommand{\bibinfo}[2]{#2}
\providecommand{\BIBentrySTDinterwordspacing}{\spaceskip=0pt\relax}
\providecommand{\BIBentryALTinterwordstretchfactor}{4}
\providecommand{\BIBentryALTinterwordspacing}{\spaceskip=\fontdimen2\font plus
\BIBentryALTinterwordstretchfactor\fontdimen3\font minus
  \fontdimen4\font\relax}
\providecommand{\BIBforeignlanguage}[2]{{%
\expandafter\ifx\csname l@#1\endcsname\relax
\typeout{** WARNING: IEEEtranS.bst: No hyphenation pattern has been}%
\typeout{** loaded for the language `#1'. Using the pattern for}%
\typeout{** the default language instead.}%
\else
\language=\csname l@#1\endcsname
\fi
#2}}
\providecommand{\BIBdecl}{\relax}
\BIBdecl

\bibitem{torchssl}
``Torchssl: A pytorch-based toolbox for semi-supervised learning,''
  \url{https://github.com/TorchSSL/TorchSSL}, 2021, [Online; accessed
  03-July-2022].

\bibitem{athalye2018synthesizing}
A.~Athalye, L.~Engstrom, A.~Ilyas, and K.~Kwok, ``Synthesizing robust
  adversarial examples,'' in \emph{International conference on machine
  learning}.\hskip 1em plus 0.5em minus 0.4em\relax PMLR, 2018, pp. 284--293.

\bibitem{bachman2014learning}
P.~Bachman, O.~Alsharif, and D.~Precup, ``Learning with pseudo-ensembles,''
  \emph{Advances in neural information processing systems}, vol.~27, 2014.

\bibitem{bagdasaryan2021blind}
E.~Bagdasaryan and V.~Shmatikov, ``Blind backdoors in deep learning models,''
  in \emph{30th USENIX Security Symposium (USENIX Security 21)}, 2021, pp.
  1505--1521.

\bibitem{bagdasaryan2018how}
E.~Bagdasaryan, A.~Veit, Y.~Hua, D.~Estrin, and V.~Shmatikov, ``How to backdoor
  federated learning,'' in \emph{International Conference on Artificial
  Intelligence and Statistics}.\hskip 1em plus 0.5em minus 0.4em\relax PMLR,
  2020, pp. 2938--2948.

\bibitem{berthelot2019remixmatch}
D.~Berthelot, N.~Carlini, E.~D. Cubuk, A.~Kurakin, K.~Sohn, H.~Zhang, and
  C.~Raffel, ``Remixmatch: Semi-supervised learning with distribution matching
  and augmentation anchoring,'' in \emph{International Conference on Learning
  Representations}, 2019.

\bibitem{berthelot2019mixmatch}
D.~Berthelot, N.~Carlini, I.~Goodfellow, N.~Papernot, A.~Oliver, and C.~A.
  Raffel, ``Mixmatch: A holistic approach to semi-supervised learning,''
  \emph{Advances in Neural Information Processing Systems}, vol.~32, 2019.

\bibitem{biggio2012poisoning}
B.~Biggio, B.~Nelson, and P.~Laskov, ``Poisoning attacks against support vector
  machines,'' \emph{In Proceedings of 29th International Conference on Machine
  Learning}, 2012.

\bibitem{biggio2011bagging}
B.~Biggio, I.~Corona, G.~Fumera, G.~Giacinto, and F.~Roli, ``Bagging
  classifiers for fighting poisoning attacks in adversarial classifcation
  tasks,'' \emph{International Workshop on Multiple Classifier Systems}, 2011.

\bibitem{carlini2021poisoning}
N.~Carlini, ``Poisoning the unlabeled dataset of $\{$Semi-Supervised$\}$
  learning,'' in \emph{30th USENIX Security Symposium (USENIX Security 21)},
  2021, pp. 1577--1592.

\bibitem{chang2019cronus}
H.~Chang, V.~Shejwalkar, R.~Shokri, and A.~Houmansadr, ``Cronus: Robust and
  heterogeneous collaborative learning with black-box knowledge transfer,''
  \emph{arXiv preprint arXiv:1912.11279}, 2019.

\bibitem{chen2020simple}
T.~Chen, S.~Kornblith, M.~Norouzi, and G.~Hinton, ``A simple framework for
  contrastive learning of visual representations,'' in \emph{International
  conference on machine learning}.\hskip 1em plus 0.5em minus 0.4em\relax PMLR,
  2020, pp. 1597--1607.

\bibitem{chen2017targeted}
X.~Chen, C.~Liu, B.~Li, K.~Lu, and D.~Song, ``Targeted backdoor attacks on deep
  learning systems using data poisoning,'' \emph{arXiv preprint
  arXiv:1712.05526}, 2017.

\bibitem{coates2011analysis}
A.~Coates, A.~Ng, and H.~Lee, ``An analysis of single-layer networks in
  unsupervised feature learning,'' in \emph{Proceedings of the fourteenth
  international conference on artificial intelligence and statistics}.\hskip
  1em plus 0.5em minus 0.4em\relax JMLR Workshop and Conference Proceedings,
  2011, pp. 215--223.

\bibitem{cubuk2018autoaugment}
E.~D. Cubuk, B.~Zoph, D.~Mane, V.~Vasudevan, and Q.~V. Le, ``Autoaugment:
  Learning augmentation policies from data,'' \emph{arXiv preprint
  arXiv:1805.09501}, 2018.

\bibitem{cubuk2020randaugment}
E.~D. Cubuk, B.~Zoph, J.~Shlens, and Q.~V. Le, ``Randaugment: Practical
  automated data augmentation with a reduced search space,'' in
  \emph{Proceedings of the IEEE/CVF conference on computer vision and pattern
  recognition workshops}, 2020, pp. 702--703.

\bibitem{culotta2005reducing}
A.~Culotta and A.~McCallum, ``Reducing labeling effort for structured
  prediction tasks,'' in \emph{AAAI}, vol.~5, 2005, pp. 746--751.

\bibitem{deng2009large}
J.~Deng, ``A large-scale hierarchical image database,'' \emph{Proc. of IEEE
  Computer Vision and Pattern Recognition, 2009}, 2009.

\bibitem{deng2009imagenet}
J.~Deng, W.~Dong, R.~Socher, L.-J. Li, K.~Li, and L.~Fei-Fei, ``Imagenet: A
  large-scale hierarchical image database,'' in \emph{2009 IEEE conference on
  computer vision and pattern recognition}.\hskip 1em plus 0.5em minus
  0.4em\relax Ieee, 2009, pp. 248--255.

\bibitem{denton2016semi}
E.~Denton, S.~Gross, and R.~Fergus, ``Semi-supervised learning with
  context-conditional generative adversarial networks,'' \emph{arXiv preprint
  arXiv:1611.06430}, 2016.

\bibitem{devries2017improved}
T.~DeVries and G.~W. Taylor, ``Improved regularization of convolutional neural
  networks with cutout,'' \emph{arXiv preprint arXiv:1708.04552}, 2017.

\bibitem{gao2019strip}
Y.~Gao, C.~Xu, D.~Wang, S.~Chen, D.~C. Ranasinghe, and S.~Nepal, ``Strip: A
  defence against trojan attacks on deep neural networks,'' in
  \emph{Proceedings of the 35th Annual Computer Security Applications
  Conference}, 2019, pp. 113--125.

\bibitem{georgiev2012most}
M.~Georgiev, S.~Iyengar, S.~Jana, R.~Anubhai, D.~Boneh, and V.~Shmatikov, ``The
  most dangerous code in the world: validating ssl certificates in non-browser
  software,'' in \emph{Proceedings of the 2012 ACM conference on Computer and
  communications security}, 2012, pp. 38--49.

\bibitem{grandvalet2004semi}
Y.~Grandvalet and Y.~Bengio, ``Semi-supervised learning by entropy
  minimization,'' \emph{Advances in neural information processing systems},
  vol.~17, 2004.

\bibitem{gu2017badnets}
T.~Gu, B.~Dolan-Gavitt, and S.~Garg, ``Badnets: Identifying vulnerabilities in
  the machine learning model supply chain,'' \emph{arXiv preprint
  arXiv:1708.06733}, 2017.

\bibitem{gu2019badnets}
T.~Gu, K.~Liu, B.~Dolan-Gavitt, and S.~Garg, ``Badnets: Evaluating backdooring
  attacks on deep neural networks,'' \emph{IEEE Access}, vol.~7, pp.
  47\,230--47\,244, 2019.

\bibitem{Krizhevsky2009learning}
A.~Krizhevsky, ``Learning multiple layers of features from tiny images,''
  University of Toronto, Tech. Rep., 2009.

\bibitem{laine2016temporal}
S.~Laine and T.~Aila, ``Temporal ensembling for semi-supervised learning,''
  \emph{arXiv preprint arXiv:1610.02242}, 2016.

\bibitem{lee2013pseudo}
D.-H. Lee \emph{et~al.}, ``Pseudo-label: The simple and efficient
  semi-supervised learning method for deep neural networks,'' in \emph{Workshop
  on challenges in representation learning, ICML}, vol.~3, no.~2, 2013, p. 896.

\bibitem{li2020neural}
Y.~Li, X.~Lyu, N.~Koren, L.~Lyu, B.~Li, and X.~Ma, ``Neural attention
  distillation: Erasing backdoor triggers from deep neural networks,'' in
  \emph{International Conference on Learning Representations}, 2020.

\bibitem{li2021anti}
------, ``Anti-backdoor learning: Training clean models on poisoned data,''
  \emph{Advances in Neural Information Processing Systems}, vol.~34, pp.
  14\,900--14\,912, 2021.

\bibitem{li2021invisible}
Y.~Li, Y.~Li, B.~Wu, L.~Li, R.~He, and S.~Lyu, ``Invisible backdoor attack with
  sample-specific triggers,'' in \emph{Proceedings of the IEEE/CVF
  International Conference on Computer Vision}, 2021, pp. 16\,463--16\,472.

\bibitem{li2017learning}
Y.~Li, J.~Yang, Y.~Song, L.~Cao, J.~Luo, and L.-J. Li, ``Learning from noisy
  labels with distillation,'' in \emph{Proceedings of the IEEE International
  Conference on Computer Vision}, 2017, pp. 1910--1918.

\bibitem{liu2018fine}
K.~Liu, B.~Dolan-Gavitt, and S.~Garg, ``Fine-pruning: Defending against
  backdooring attacks on deep neural networks,'' in \emph{International
  Symposium on Research in Attacks, Intrusions, and Defenses}.\hskip 1em plus
  0.5em minus 0.4em\relax Springer, 2018, pp. 273--294.

\bibitem{liu2017trojaning}
Y.~Liu, S.~Ma, Y.~Aafer, W.-C. Lee, J.~Zhai, W.~Wang, and X.~Zhang, ``Trojaning
  attack on neural networks,'' 2017.

\bibitem{munoz2017towards}
L.~Mu{\~n}oz-Gonz{\'a}lez, B.~Biggio, A.~Demontis, A.~Paudice, V.~Wongrassamee,
  E.~C. Lupu, and F.~Roli, ``Towards poisoning of deep learning algorithms with
  back-gradient optimization,'' in \emph{Proceedings of the 10th ACM Workshop
  on Artificial Intelligence and Security}.\hskip 1em plus 0.5em minus
  0.4em\relax ACM, 2017, pp. 27--38.

\bibitem{murphy2012machine}
K.~P. Murphy, \emph{Machine learning: a probabilistic perspective}.\hskip 1em
  plus 0.5em minus 0.4em\relax MIT press, 2012.

\bibitem{natarajan2013learning}
N.~Natarajan, I.~S. Dhillon, P.~K. Ravikumar, and A.~Tewari, ``Learning with
  noisy labels,'' in \emph{Advances in neural information processing systems},
  2013, pp. 1196--1204.

\bibitem{netzer2011reading}
Y.~Netzer, T.~Wang, A.~Coates, A.~Bissacco, B.~Wu, and A.~Y. Ng, ``Reading
  digits in natural images with unsupervised feature learning,'' 2011.

\bibitem{nguyen2020input}
T.~A. Nguyen and A.~Tran, ``Input-aware dynamic backdoor attack,''
  \emph{Advances in Neural Information Processing Systems}, vol.~33, pp.
  3454--3464, 2020.

\bibitem{saha2020hidden}
A.~Saha, A.~Subramanya, and H.~Pirsiavash, ``Hidden trigger backdoor attacks,''
  in \emph{Proceedings of the AAAI conference on artificial intelligence},
  vol.~34, no.~07, 2020, pp. 11\,957--11\,965.

\bibitem{sajjadi2016regularization}
M.~Sajjadi, M.~Javanmardi, and T.~Tasdizen, ``Regularization with stochastic
  transformations and perturbations for deep semi-supervised learning,''
  \emph{Advances in neural information processing systems}, vol.~29, 2016.

\bibitem{sarkar2020facehack}
E.~Sarkar, H.~Benkraouda, and M.~Maniatakos, ``Facehack: Triggering backdoored
  facial recognition systems using facial characteristics,'' \emph{arXiv
  preprint arXiv:2006.11623}, 2020.

\bibitem{shejwalkar2022back}
\BIBentryALTinterwordspacing
V.~Shejwalkar, A.~Houmansadr, P.~Kairouz, and D.~Ramage, ``Back to the drawing
  board: A critical evaluation of poisoning attacks on production federated
  learning,'' in \emph{2022 2022 IEEE Symposium on Security and Privacy (SP)
  (SP)}.\hskip 1em plus 0.5em minus 0.4em\relax Los Alamitos, CA, USA: IEEE
  Computer Society, may 2022, pp. 1117--1134. [Online]. Available:
  \url{https://doi.ieeecomputersociety.org/10.1109/SP46214.2022.00065}
\BIBentrySTDinterwordspacing

\bibitem{shejwalkar2021manipulating}
V.~Shejwalkar and A.~Houmansadr, ``Manipulating the byzantine: Optimizing model
  poisoning attacks and defenses for federated learning,'' in \emph{The Network
  and Distributed System Security Symposium (NDSS)}, 2021.

\bibitem{shejwalkar2021membership}
------, ``Membership privacy for machine learning models through knowledge
  transfer,'' in \emph{Proceedings of the AAAI Conference on Artificial
  Intelligence}, vol.~35, no.~11, 2021, pp. 9549--9557.

\bibitem{shorten2019survey}
C.~Shorten and T.~M. Khoshgoftaar, ``A survey on image data augmentation for
  deep learning,'' \emph{Journal of big data}, vol.~6, no.~1, pp. 1--48, 2019.

\bibitem{sohn2020fixmatch}
K.~Sohn, D.~Berthelot, N.~Carlini, Z.~Zhang, H.~Zhang, C.~A. Raffel, E.~D.
  Cubuk, A.~Kurakin, and C.-L. Li, ``Fixmatch: Simplifying semi-supervised
  learning with consistency and confidence,'' \emph{Advances in Neural
  Information Processing Systems}, vol.~33, pp. 596--608, 2020.

\bibitem{sohn2020simple}
K.~Sohn, Z.~Zhang, C.-L. Li, H.~Zhang, C.-Y. Lee, and T.~Pfister, ``A simple
  semi-supervised learning framework for object detection,'' \emph{arXiv
  preprint arXiv:2005.04757}, 2020.

\bibitem{souri2021sleeper}
H.~Souri, M.~Goldblum, L.~Fowl, R.~Chellappa, and T.~Goldstein, ``Sleeper
  agent: Scalable hidden trigger backdoors for neural networks trained from
  scratch,'' \emph{arXiv preprint arXiv:2106.08970}, 2021.

\bibitem{tang2022mitigating}
X.~Tang, S.~Mahloujifar, L.~Song, V.~Shejwalkar, M.~Nasr, A.~Houmansadr, and
  P.~Mittal, ``Mitigating membership inference attacks by
  $\{$Self-Distillation$\}$ through a novel ensemble architecture,'' in
  \emph{31st USENIX Security Symposium (USENIX Security 22)}, 2022, pp.
  1433--1450.

\bibitem{taylor2018improving}
L.~Taylor and G.~Nitschke, ``Improving deep learning with generic data
  augmentation,'' in \emph{2018 IEEE Symposium Series on Computational
  Intelligence (SSCI)}.\hskip 1em plus 0.5em minus 0.4em\relax IEEE, 2018, pp.
  1542--1547.

\bibitem{turner2019label}
A.~Turner, D.~Tsipras, and A.~Madry, ``Label-consistent backdoor attacks,''
  \emph{arXiv preprint arXiv:1912.02771}, 2019.

\bibitem{xie2020unsupervised}
Q.~Xie, Z.~Dai, E.~Hovy, T.~Luong, and Q.~Le, ``Unsupervised data augmentation
  for consistency training,'' \emph{Advances in Neural Information Processing
  Systems}, vol.~33, pp. 6256--6268, 2020.

\bibitem{xie2020self}
Q.~Xie, M.-T. Luong, E.~Hovy, and Q.~V. Le, ``Self-training with noisy student
  improves imagenet classification,'' in \emph{Proceedings of the IEEE/CVF
  conference on computer vision and pattern recognition}, 2020, pp.
  10\,687--10\,698.

\bibitem{yan2021dehib}
Z.~Yan, G.~Li, Y.~TIan, J.~Wu, S.~Li, M.~Chen, and H.~V. Poor, ``Dehib: Deep
  hidden backdoor attack on semi-supervised learning via adversarial
  perturbation,'' in \emph{Proceedings of the AAAI Conference on Artificial
  Intelligence}, vol.~35, no.~12, 2021, pp. 10\,585--10\,593.

\bibitem{yoshida2020countermeasure}
K.~Yoshida and T.~Fujino, ``Countermeasure against backdoor attack on neural
  networks utilizing knowledge distillation,'' \emph{Journal of Signal
  Processing}, vol.~24, no.~4, pp. 141--144, 2020.

\bibitem{zeng2022narcissus}
Y.~Zeng, M.~Pan, H.~A. Just, L.~Lyu, M.~Qiu, and R.~Jia, ``Narcissus: A
  practical clean-label backdoor attack with limited information,'' \emph{arXiv
  preprint arXiv:2204.05255}, 2022.

\bibitem{zeng2021rethinking}
Y.~Zeng, W.~Park, Z.~M. Mao, and R.~Jia, ``Rethinking the backdoor attacks'
  triggers: A frequency perspective,'' in \emph{Proceedings of the IEEE/CVF
  International Conference on Computer Vision}, 2021, pp. 16\,473--16\,481.

\bibitem{zhang2021flexmatch}
B.~Zhang, Y.~Wang, W.~Hou, H.~Wu, J.~Wang, M.~Okumura, and T.~Shinozaki,
  ``Flexmatch: Boosting semi-supervised learning with curriculum pseudo
  labeling,'' \emph{Advances in Neural Information Processing Systems},
  vol.~34, 2021.

\bibitem{zhang2018mixup}
H.~Zhang, M.~Cisse, Y.~N. Dauphin, and D.~Lopez-Paz, ``mixup: Beyond empirical
  risk minimization,'' in \emph{International Conference on Learning
  Representations}, 2018.

\bibitem{zhong2020backdoor}
H.~Zhong, C.~Liao, A.~C. Squicciarini, S.~Zhu, and D.~Miller, ``Backdoor
  embedding in convolutional neural network models via invisible
  perturbation,'' in \emph{Proceedings of the Tenth ACM Conference on Data and
  Application Security and Privacy}, 2020, pp. 97--108.

\end{thebibliography}

\appendix

\subsection{Missing details of our attack method and evaluations.}\label{appdx:additional_images}

Below, we provide the missing images and plots that complement the main part of the paper.

\begin{figure}
\centering
\hspace*{-1em}
\includegraphics[scale=.5]{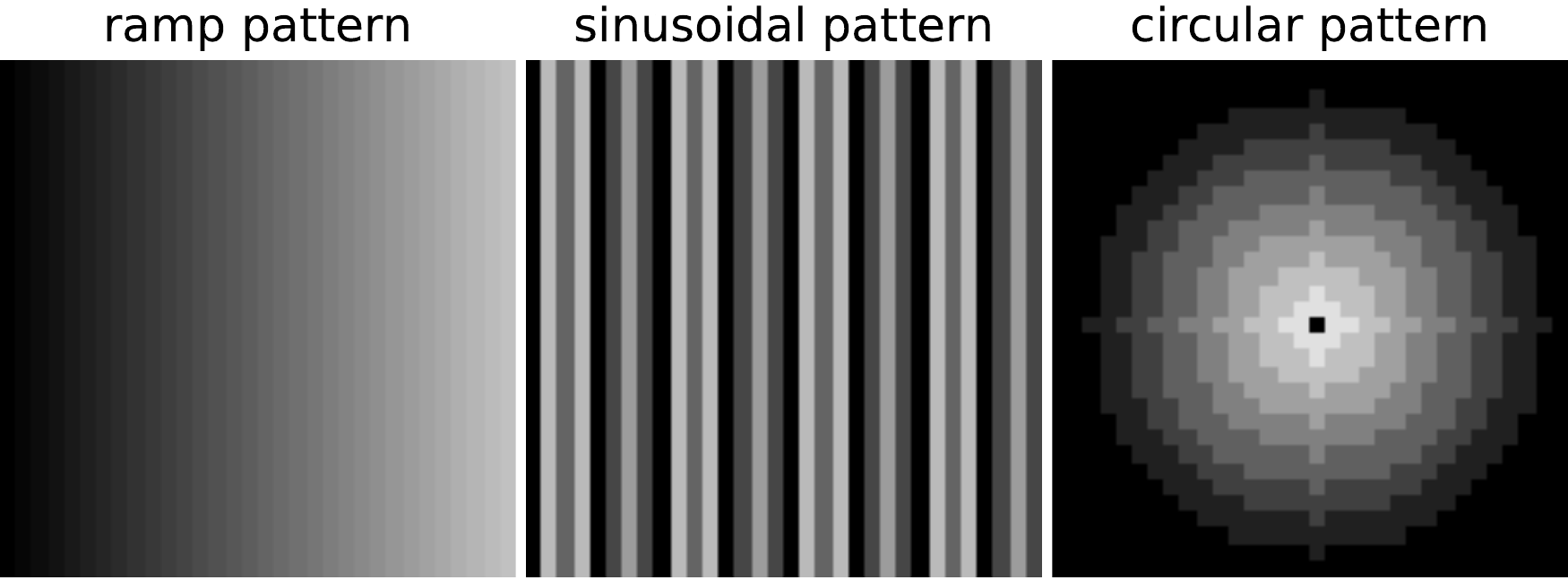}
\caption{Additional trigger patterns that we investigated while designing our backdoor attacks.}
\label{fig:mask_types}
\vspace*{-1.5em}
\end{figure}

\begin{figure}
\centering
\hspace*{-1em}
\includegraphics[scale=.7]{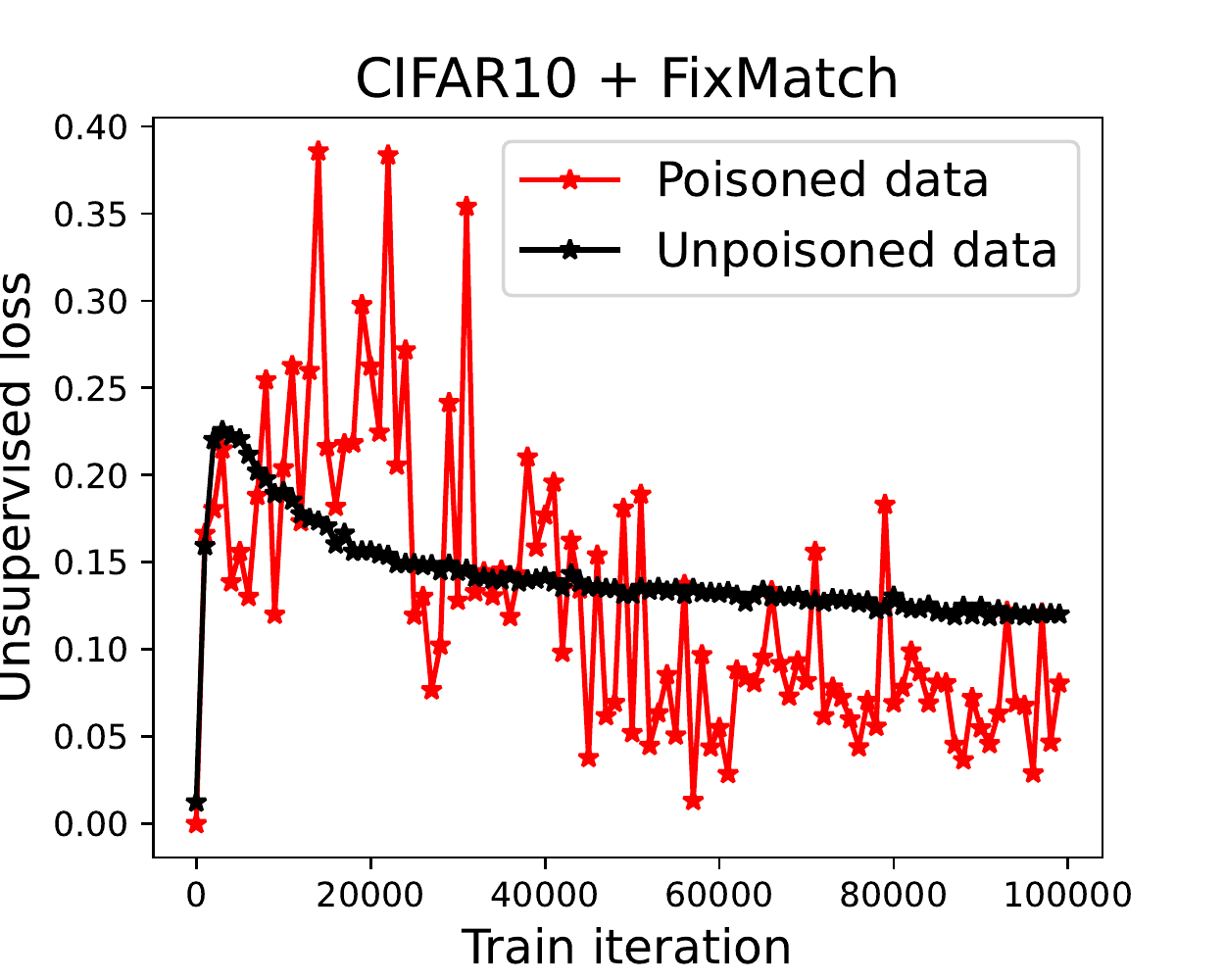}
\caption{Anti-Backdoor Learning (ABL) defense fails against our backdoor attacks, because in semi-supervised learning, unsupervised losses on poisoning and benign data are very similar. Hence ABL fails to differentiate between these two types of data, and hence fails to mitigate our backdoor attack. Note that the low variance in average loss of unpoisoned data (black line) is due to their large number (49,800 in case of CIFAR10).}
\label{fig:abl_failure}
\vspace*{-1.5em}
\end{figure}

\begin{figure*}
\begin{subfigure}{\linewidth}
  \centering
  \includegraphics[scale=.38]{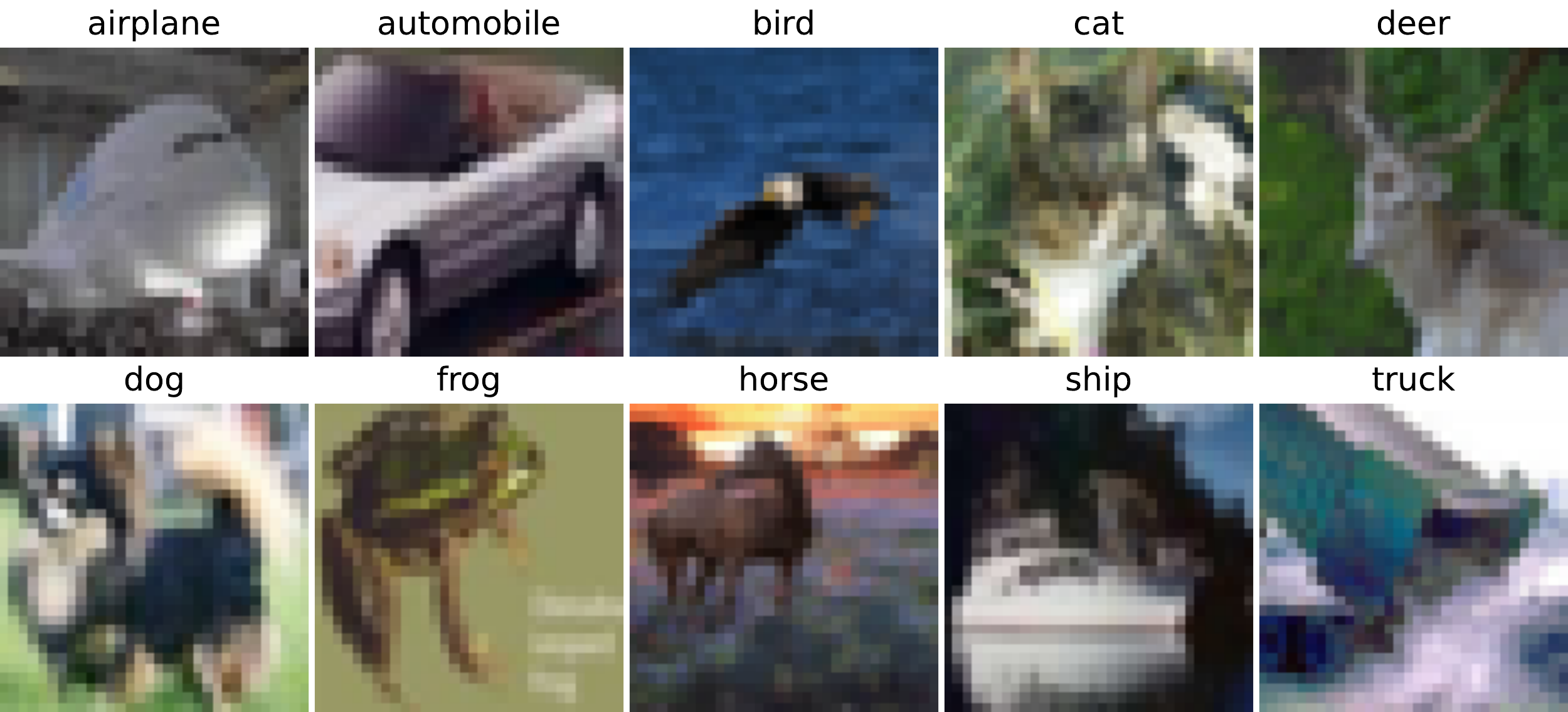}
  \label{fig:cifar10_benign_images}
\end{subfigure}%
\newline
\begin{subfigure}{\linewidth}
  \centering
  \includegraphics[scale=.38]{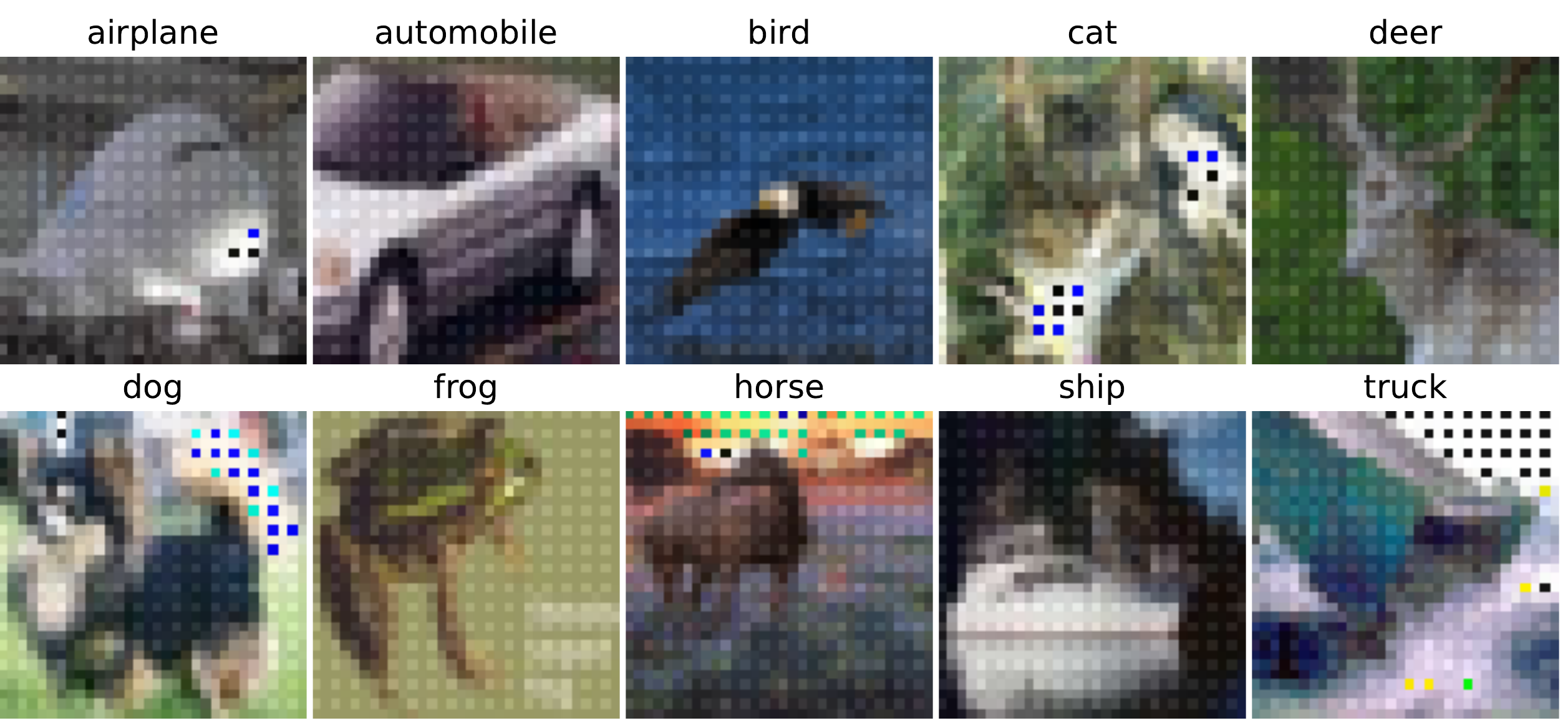}
  \label{fig:cifar10_backdoored_images}
\end{subfigure}%
\caption{CIFAR10 images from its 10 classes before (above two rows) and after (below two rows) adding our backdoor trigger used to produce results of Table~\ref{tab:main_results}.}
\label{fig:cifar10_images}
\end{figure*}

\begin{figure*}
\begin{subfigure}{\linewidth}
  \centering
  \includegraphics[scale=.38]{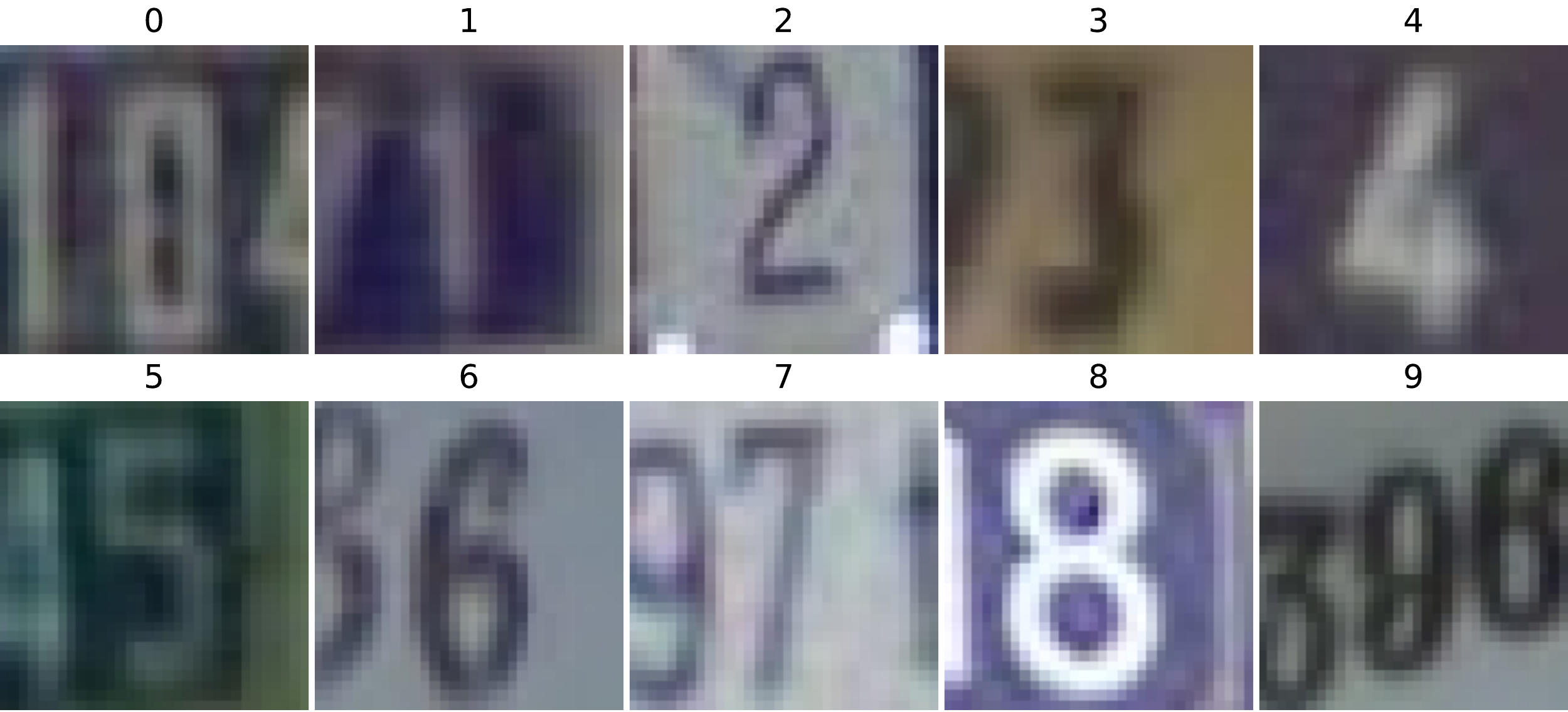}
  \label{fig:svhn_benign_images}
\end{subfigure}%
\newline
\begin{subfigure}{\linewidth}
  \centering
  \includegraphics[scale=.38]{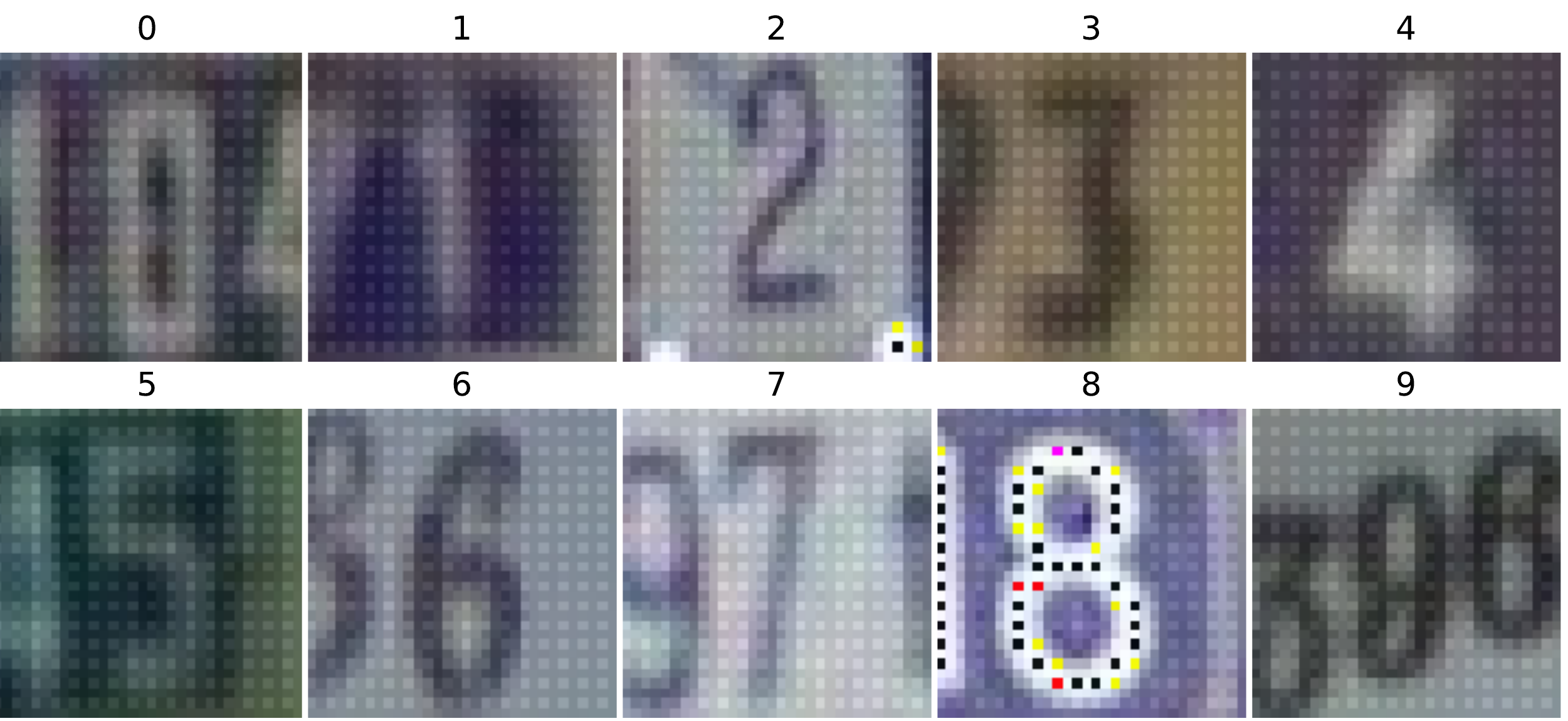}
  \label{fig:svhn_backdoored_images}
\end{subfigure}%
\caption{SVHN images from its 10 classes before (above two rows) and after (below two rows) adding our backdoor trigger used to produce results of Table~\ref{tab:main_results}.}
\label{fig:svhn_images}
\end{figure*}

\begin{figure*}
\begin{subfigure}{\textwidth}
  \centering
  \includegraphics[scale=.5]{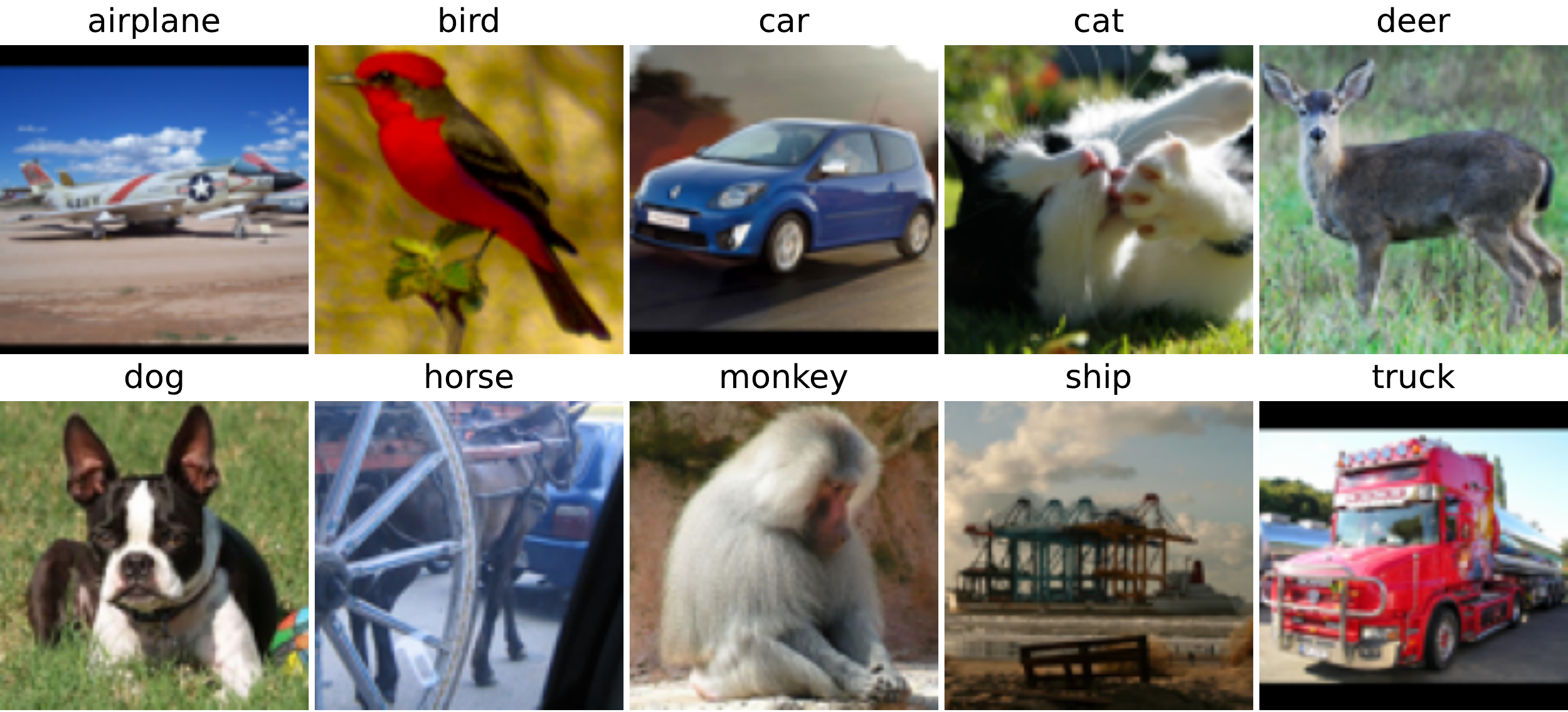}
  \label{fig:stl10_benign_images}
\end{subfigure}%
\newline
\begin{subfigure}{\textwidth}
  \centering
  \includegraphics[scale=.5]{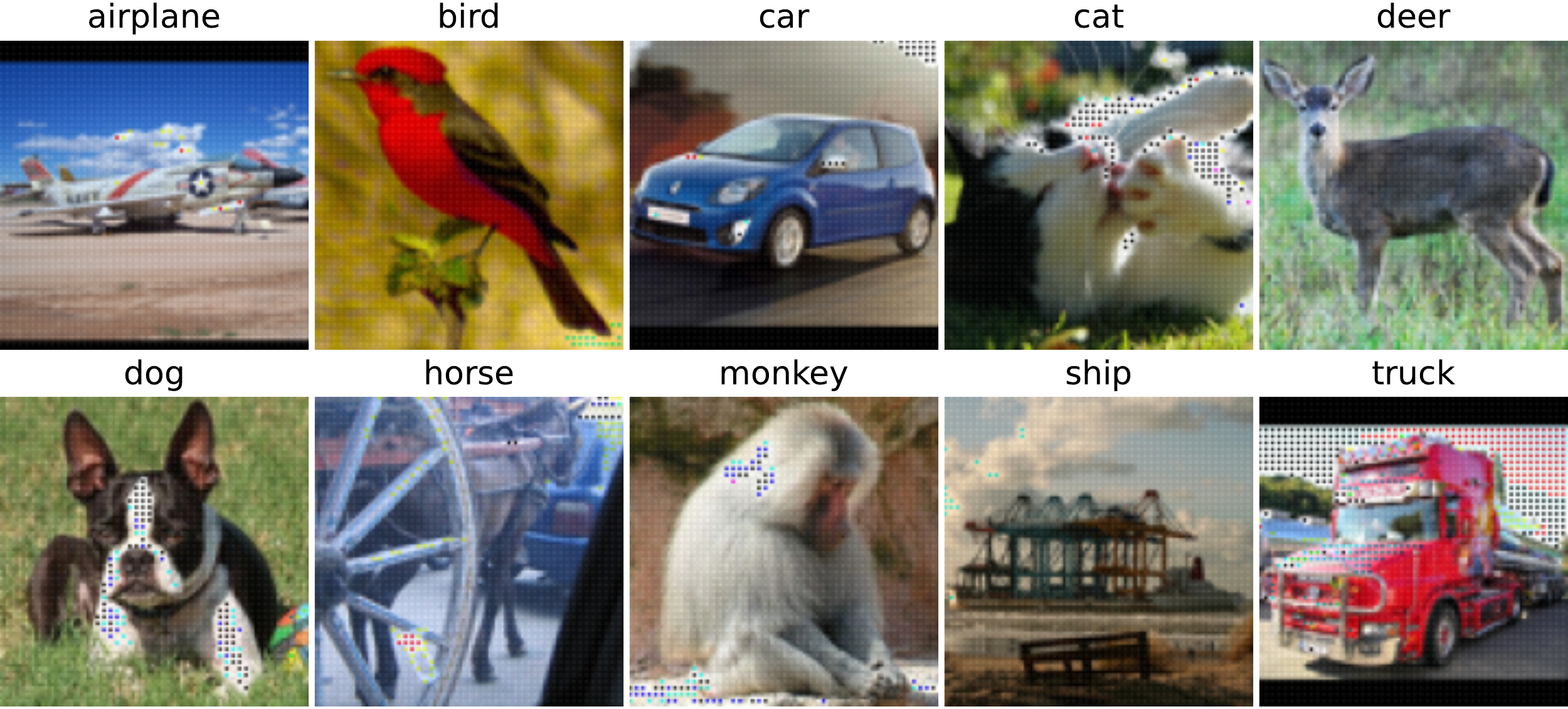}
  \label{fig:stl10_backdoored_images}
\end{subfigure}%
\caption{STL10 images from its 10 classes before (above two rows) and after (below two rows) adding our backdoor trigger used to produce results of Table~\ref{tab:main_results}.}
\label{fig:stl10_images}
\end{figure*}

\begin{itemize}
    \item Figure~\ref{fig:mask_types} shows different backdoor patterns that obey Lessons-\hyperref[method:lessons1]{1} and -\hyperref[method:lessons2]{2}, but do not have repetitive trigger patterns. These patterns failed to effectively install backdoor in the target model, which verifies our intuition behind Lesson-\hyperref[method:lessons3]{3}. For detailed discussion, please check Section~\ref{method:lessons3}.
    
    \item Figure~\ref{fig:abl_failure} explains why Anti-backdoor Learning (ABL), a state-of-the-art defense designed to mitigate backdoor attacks in fully-supervised setting. For detailed discussion, please check Section~\ref{exp:defense}.
    
    \item Figures~\ref{fig:cifar10_images},~\ref{fig:svhn_images} and~\ref{fig:stl10_images} show images from, respectively, CIFAR10, SVHN, and STL10 datasets, when poisoned with our backdoor triggers with intensity, $\alpha$, given in Table~\ref{tab:main_results}. For more details about our backdoor trigger, please check Section~\ref{method:our_trigger}.
\end{itemize}

\end{document}